\documentclass[aps,twocolumn,showpacs,superscriptaddress,eqsecnum]{revtex4}

\usepackage{graphicx}
\usepackage{amsfonts}
\usepackage[figuresright]{rotating}  
\usepackage{amssymb}
\usepackage{amsmath}
\usepackage{subfigure}
\usepackage{multirow}
\usepackage{tabularx}

\def\LSCO{La$_{2-x}$Sr$_x$CuO$_4$}

\def\LBCO{La$_{2-x}$Ba$_x$CuO$_4$}
\def\YBCO{YBa$_2$Cu$_3$O$_{6+y}$}

\def\BSCCO{Bi$_2$Sr$_2$CaCu$_2$O$_{8+\delta}$}
\def\C60{A$_x$C$_{60}$}

\def\LNSCO{La$_{1.6-x}$Nd$_{0.4}$Sr$_x$CuO$_{4}$}

\def\HgCu3{HgCa$_2$Cu$_3$O$_{8+y}$}
\def\HgCu4{HgBa$_2$Ca$_3$Cu$_4$O$_{10+y}$}
\def\TlCu{Tl$_2$Ba$_2$CuO$_{6+\delta}$}
\def\TlCu3{Tl$_2$Ba$_2$Ca$_2$Cu$_3$O$_{10+y}$}
\def\TlCu4{Tl$_2$Ba$_2$Ca$_3$Cu$_4$O$_{12+y}$}

\def\BiCu3{Bi$_2$Sr$_2$Ca$_{2}$Cu$_3$O$_y$}

\def\8LSCO{La$_{1.88}$Sr$_{.12}$CuO$_4$}
\def\110LNSCO{La$_{1.5}$Nd$_{0.4}$Sr$_{0.1}$CuO$_{4}$}
\def\stage4LCO{La$_{2}$CuO$_{4+\delta}$}
\def\Y248{YBa$_2$Cu$_4$O$_8$}

\def\hty{high temperature superconductivity}
\def\hts{high temperature superconductors}
\def\HTS{High temperature superconductors}

\def\NbSe2{NbSe$_2$}
\def\TaSe2{TaSe$_2$}
\def\TiSe2{TiSe$_2$}
\def\NaCoOH2O{Na$_{0.3}$CoO$_{2y}$H$_2$O}
\def\MgB2{MgB${}_2$}
\def\doubleRu{Sr$_3$Ru$_2$O$_7$}

\def\avg#1{\langle#1\rangle}

\def\tr{\mbox{tr}}
\def\nn{\nonumber}

\begin{document}

\title{Fluctuating Stripes in Strongly Correlated Electron Systems and the  Nematic-Smectic Quantum Phase Transition}
\author{Kai Sun}
\affiliation{Department of Physics, University of Illinois at Urbana-Champaign, Urbana, Illinois 61801-3080, USA}
\author{Benjamin M. Fregoso}
\affiliation{Department of Physics, University of Illinois at Urbana-Champaign, Urbana, Illinois 61801-3080, USA}
\author{Michael J. Lawler}
\affiliation{Department of Physics, University of Toronto,  Toronto, Ontario M5S 1A7, Canada}
\author{Eduardo Fradkin}
\affiliation{Department of Physics, University of Illinois at Urbana-Champaign, Urbana, Illinois 61801-3080, USA}

\begin{abstract}
We discuss the quantum phase transition between a quantum nematic 
metallic state to an electron metallic smectic state in terms of
an order-parameter theory coupled to fermionic quasiparticles.
Both commensurate and incommensurate smectic (or stripe) cases are studied.  
Close to the quantum critical point (QCP), the spectrum of fluctuations of the nematic phase has low-energy ``fluctuating stripes''.  
We study the quantum critical behavior and find evidence that, contrary to the classical case, the 
gauge-type of coupling between the nematic and smectic is irrelevant at 
this QCP. The collective modes of the electron smectic (or stripe) phase are also investigated.
The effects of the low-energy bosonic modes on the fermionic quasiparticles 
are studied perturbatively, for both a model with full rotational symmetry and for a system with an underlying lattice, which has a discrete point group symmetry. We find that at the nematic-smectic critical point, due to the critical 
smectic fluctuations, the dynamics of the fermionic quasiparticles near 
several points on the Fermi surface, around which it is reconstructed,
are not governed by a Landau Fermi liquid theory. On the other hand, the quasiparticles in the smectic phase 
exhibit Fermi liquid behavior.
We also present a detailed analysis of the dynamical susceptibilities in the electron nematic phase close to this QCP (the fluctuating stripe regime) and in the electronic smectic phase.
\end{abstract}
\date{\today}
\maketitle 

\section{Introduction}
\label{sec:intro}

The discovery of the {\hts} in the quasi-two-dimensional copper-oxide materials in the late 1980s, 
and of novel correlated phases in other complex oxides,  
has brought to the forefront the problem of  the physics of strongly correlated electron systems. 
To this date the understanding of the behavior of these systems remains one of the main open and challenging problems in condensed matter physics. 
The central conundrum in this field is the fact that these strongly coupled electron systems 
are best regarded as doped Mott insulators for which both the band theory of metals and the Landau theory of the Fermi liquid (FL) fail.

One characteristic feature of the physics of doped Mott insulators is their inherent tendency to electronic phase separation, frustrated by the effects of Coulomb interactions.\cite{emery93,kivelson-1994} The ground states resulting from these competing tendencies typically break the translation invariance and/or the point group symmetry of the underlying lattice. From a symmetry point of view, the ground states of doped Mott insulators are charge-ordered phases, which share many similarities with classical liquid crystals, and should be regarded as electronic liquid crystal phases.\cite{Kivelson1998} However,
unlike classical liquid crystals, electronic liquid crystals are strongly quantum mechanical states whose transport properties range from insulating to  metallic and even superconducting. In contrast with classical liquid crystals, whose ordered phases represent the spontaneous breaking of the continuous translation and rotational symmetry of space\cite{DeGennes1993,Chaikin98}, the electronic liquid crystal phases of strongly correlated systems are sensitive to the effects of the underlying lattice and the symmetry breaking patterns involve the point and space groups, as well as to disorder. More complex  ordered states, involving simultaneously charge and spin degrees of freedom, may also arise.\cite{wu-sun-fradkin-zhang-2007}
 
 The sequence of quantum phase transitions described above,  electron crystal $\rightarrow$ smectic (stripe) $\rightarrow$ nematic $\rightarrow$ isotropic fluid, representing the progressive restoration of symmetry, is natural from a strong correlation perspective. Indeed, the electron crystal state(s) are naturally  insulating (much as in the case of a Wigner crystal), the smectic or stripe phases are either anisotropic metals or superconductors, and the charged isotropic fluids are either metallic or superconducting. While the isotropic metallic phase is essentially a FL (albeit with strongly renormalized parameters), the nematic and smectic metallic phases have a strong tendency to show non-FL character. Indeed, much of the theoretical description of the stripe or smectic phases is usually based on  a quasi-one-dimensional  analysis, which makes explicit use of this strong correlation physics. Such approaches give a good description of this state deep inside this phase and at energies high compared to a ``dimensional crossover'' scale below which the state is fully two-dimensional (and strongly anisotropic)\cite{emery97,carlson00,Emery00,Vishwanath01,granath01,arrigoni-fradkin-kivelson-2004,carlson04}. Stripe phases (insulating, metallic, and superconducting) have been found in mean-field studies of generalized two-dimensional Hubbard and t-J models\cite{zaanen-1989,machida-1989,kato-1990,poilblanc-1989,schulz-1990,vojta99,vojta-2000,park-2001,sachdev03,lorenzana02,lorenzana04,himeda-2002,raczkowski-2007}.
 
 The same pattern of quantum phase transitions can also be considered in reverse order, with a weak coupling perspective, as a sequence of symmetry breaking phase transitions beginning from the isotropic metal: FL $\rightarrow$ electron nematic $\rightarrow$ electron smectic $\rightarrow$ insulating electron crystal. 
In this case, one begins with a uniform isotropic metal, well described at low energies by the Landau theory of the FL,  with well-defined quasiparticles and a Fermi surface (FS), and considers possible instabilities of the isotropic fluid into a nematic (or hexatic and other such states), as well as phase transitions into various possible charge-density-wave (CDW) phases. The unidirectional CDW-ordered states are the weak coupling analog of the smectic (or stripe) phases, and have the same order parameters as they break the same symmetries. The main difference between a CDW and a smectic resides in the fact that while the CDW arises as a weak coupling (infinitesimal) instability of a FL in which parts of the FS are gapped\cite{mcmillan75} (which requires the existence of a FS with sharp quasiparticles),  the stripe phases do not require such description. While a CDW phase at high energies is essentially a FL, the high-energy regime of a stripe phase is a quasi-one-dimensional Luttinger liquid.\cite{carlson00,Emery00} A direct quantum phase transition from a FL to a CDW phase is, naturally, possible and this quantum phase transition has been studied in some detail,\cite{Altshuler1995,Chubukov2005} as well as to a metallic spin-density wave (SDW)\cite{vekhter-2004,lohneysen-2007}.
 
The weak coupling description of an electron nematic phase uses a Pomeranchuk instability of a Fermi liquid state\cite{pomeranchuk58}. Oganesyan, Kivelson, and Fradkin\cite{Oganesyan2001} showed that the nematic quantum phase transition is  a quadrupolar instability of the FS, and gave a characterization of the properties of the nematic Fermi fluid in a continuum model. An electron nematic quantum phase transition has also been found in lattice models\cite{Halboth-2000,Metzner-2003,dellanna-2006}, which show, however, a strong tendency to exhibit a first-order quantum phase transition\cite{kee-2003,Khavkine-2004,yamase-2005}. Pomeranchuk instabilities in the Landau theory of the FL have also shown the existence of an electron nematic transition\cite{Nilsson05,wolfle-2007}.  Perturbative renormalization group analysis of the stability of the FL in Hubbard-type models\cite{honerkamp-2002}, as well as high-temperature expansions\cite{pryadko-2004}, has also shown that in such models there is a strong tendency to a nematic state. An electron nematic state was shown to be the exact ground state in the strong coupling limit of the Emery model of the copper oxides at low hole doping\cite{kivelson04}.
 
 The upshot of the work on the electron nematic quantum phase transition is that, at the QCP (if the transition is continuous) and in the nematic phase (in the continuum) the electron quasiparticle essentially no longer exists as an asymptotically stable state at low energies, except along symmetry determined directions in the ordered phase. A full solution of this QCP by bosonization methods has confirmed these results, which were gleaned from mean-field theory, and have also provided strong evidence for local quantum criticality at this QCP \cite{lawler-2006,lawler-2007}.

 In this paper we will be interested in the quantum phase transition from an electron nematic phase to a charge stripe phase, a unidirectional CDW.
For simplicity we will not consider here the spin channel, which plays an important role in many systems. We will only consider the simpler case of unidirectional order. Extensions to the more general case of multidirectional order are straightforward. Here we develop a quantum-mechanical version of the nematic-smectic transition in a metallic system. This is a quantum-mechanical version of the McMillan-deGennes theory for the quantum phase transition from a metallic nematic phase to a metallic smectic (or CDW) phase. The construction of such a generalization of the McMillan-deGennes theory is the main purpose of this paper. 

As it is discussed in detail in subsequent sections, here we will follow the ``weak-coupling'' sequence of quantum phase transitions described above, beginning with the transition from a FL to an electron nematic, and from the latter to a stripe or unidirectional CDW state. The main advantages of this approach are that it allows to address the fate of the electronic quasiparticles and non-Fermi liquid behaviors as the correlations that give rise to these electronic liquid crystal phases develop, as well as to study the quantum critical behavior following the standard Hertz-Millis approach \cite{Hertz1976,Millis1993,Sachdev1999}. However, the main disadvantage is that this approach does not do justice to the physics of strong correlation. For this reason, in spite of the important insights that are gained through this line of analysis,  this approach cannot explain  the physics of the ``strange metal'' regime observed in the ``normal state'' of high $T_c$ superconductors where non-Fermi liquid effects are widely reported. To do that would require studying this problem as a sequence of quantum melting  transitions. An important first step in this direction has been made by Cvetkovic and coworkers \cite{cvetkovic-2006,cvetkovic-2007,cvetkovic-annphys} who have studied a purely bosonic model of such quantum melting. The inclusion of fermionic degrees of freedom in this strong coupling approach is an interesting but challenging open problem. 

 We have both conceptual and phenomenological motivations for considering this problem. At the conceptual level the main question is to develop a theory of the quantum critical behavior at the electron nematic-smectic phase transition, and of the low-energy physics of both phases near quantum criticality. Although the static properties are
the same as in the classical theory (as required by symmetry) the quantum dynamics changes the physics substantially.
 Thus, physical properties, which determine the transport properties and the fermion spectral function, cannot be gleaned from the classical problem. Provided that the quantum phase transition is continuous or, at most weakly first order, the low-energy fluctuations in one phase (say the nematic metal) must reflect the character of the nearby ordered stripe phase. In other words, under these assumptions, as the quantum phase transition is approached  the metallic nematic phase behaves as a state with ``fluctuating stripes''. The ample experimental evidence in {\hts} for ``fluctuating stripe order'' should be interpreted instead as evidence of a nematic phase proximate to a quantum phase transition to a stripe (or smectic)-ordered state\cite{Kivelson2003}. 
 
\section{Summary of Results}
\label{sec:summary}
 
 In this work we follow a phenomenological approach to study the quantum phase transition between an electronic nematic state and electronic smectic state. We postulate the existence of both an electron nematic and a smectic phases with a possible direct phase transition between them. This physics will be represented by an effective field theory involving the nematic and CDW order parameters. The static part of the effective action of the order-parameter theory has the same form as in the classical theory of the nematic-smectic transition, the  McMillan-deGennes theory. 
We will assume that aside from the effects of the coupling to the fermionic quasiparticles, this effective field theory is analytic in the order parameters and their derivatives as this dependence is determined by local physics. As shown below, this assumption implies
a dynamical quantum critical exponent $z=1$.

The fermionic quasiparticles couple to the nematic and smectic (CDW) order parameters in their natural symmetry-dictated way. The fermions will be assumed to be a normal FL, with well-defined quasiparticles and a FS. Thus, we will not attempt to explain why the phase transition exists, which requires a microscopic theory,  but rather describe its character. One of our  most important results is that this theory gives a description of a phase with fluctuating stripe (smectic) order, of much interest in current experiments. The  effective theory that we consider also allows for a possible direct transition between the normal and isotropic FL state and a CDW phase, without going through an intermediate nematic phase, as in the direct transition between a FL and a CDW state, discussed by Altshuler, Ioffe, and Millis\cite{Altshuler1995}. Thus, the theory we present here actually describes the behavior of a FL in the vicinity of a possible bicritical point which, as we shall see, is not directly accessible. 

\begin{widetext}
\begin{center}
\begin{table}[h]
\begin{tabular}{|c|c|c|c|c|c|c|c|}
\hline
\multirow{4}{*}{}&\multirow{4}{*}{Nematic} &\multicolumn{4}{|c|}{Smectic Mode at the Electronic Nematic-Smectic QCP} 
&\multicolumn{2}{|c|}
{Smectic} 
\\ \cline{3-8}
& & \multirow{3}{*}{$Q_S<2 k_F$} & $Q_S=2k_F$ & $Q_S=2k_F$ &inflection&continuous&discrete
\\ 
& & &incommensurate&commensurate&point&rotational&rotational
\\
& & &&&&symmetry&symmetry
\\
\hline
Anisotropic Scaling & \multirow{2}{*}{$1:1:3$} & \multirow{2}{*}{$1:1:2$} & \multirow{2}{*}{$1:2:3$}
& \multirow{2}{*}{$1:2:2$} &  \multirow{2}{*}{$1:3:3$} & \multirow{2}{*}{$1:2:2$}
& \multirow{2}{*}{$1:1:1$}
\\
$[q_x]:[q_y]:[\omega]$ &&&&&&& 
\\
\hline
Non-analyticity & & & $\Phi^{5/2}$ & $\Phi^{5/2}$ & $\Phi^{9/4}$ & &
\\
\hline
Gaussian Fixed Point & Stable & Stable & Unstable / First Order & Stable & Stable & Stable & Stable
\\
\hline
\multirow{2}{*}{$\Sigma''(k_F,\omega)$} &\multirow{2}{*}{$|\omega|^{2/3}$}&\multirow{2}{*}{$|\omega|^{1/2}$}&\multirow{2}{*}{?}&\multirow{2}{*}{$|\omega|$}&\multirow{2}{*}{$|\omega|^{13/12}$}&$\omega^2 \displaystyle{\log |\omega|}$&$\omega^2 \displaystyle{\log |\omega|}$ \\
&&&&&&or $|\omega|^{3/2}$& or $\omega^2$\\
\hline
\end{tabular}
\caption{Summary of results. See the text for a detailed explanation.}
\label{table:summary}
\end{table}
\end{center}
\end{widetext}

The main results of our theory are summarized in Table \ref{table:summary}.  In Sec. \ref{sec:expt} we discuss the current experimental status of electronic liquid crystal phases in a number of different materials. In Sec. \ref{sec:classical} we set up the order parameter theory for the electronic liquid crystal
phases based on symmetry and analyticity.  The 
static part of this phenomenological theory is (as it should be) similar to its 
classical counterpart, but we add proper dynamics to  describe the 
quantum fluctuations. We next  couple the order parameter theory to the fermionic quasiparticles, in Sec.
\ref{sec:effect_fermions}. The coupling between the fermionic quasiparticles and the order parameters is completely determined by symmetry. This is a standard approach to study quantum phase transitions in metallic systems \cite{Sachdev1999}. It is a consistent scheme for the study of the quantum phase transition provided the effective dimension $d+z$ is close to 4 (here $d$ is the dimensionality of space).  Several different non-analytic dependences on the order parameters in the effective action appear as a consequence of their coupling  to the fermions. We show that these nonanalytic dynamical terms 
dominate over the dynamics prescribed phenomenologically.
Hence, the dynamics of fermionic liquid crystal phase is very different
from that of the simple phenomenological theory. We present a detailed analysis of the behavior of the dynamical susceptibilities in both  phases and at the QCP.

The nematic-smectic QCP is studied in Sec.
\ref{sec:N_S_critical}. In classical liquid crystals, the Goldstone mode of the nematic phase
plays a very important role at the nematic-smectic transition. There, this relevant coupling drives the transition weakly first order through a fluctuation-induced first order transition\cite{Halperin1974}. However, in the 
case of the electronic liquid crystals, we find that the coupling between the nematic Goldstone 
mode and the smectic field is actually irrelevant at the electronic nematic-smectic QCP. Therefore, these two modes can be 
treated separately, as they are weakly coupled to each other. Several different nematic-smectic critical theories
are studied, depending on the relation between the magnitude of the ordering wave 
vector of the CDW, $Q_S$, and the Fermi wave vector, $k_F$.
For $Q_S<2 k_F$ (Fig. \ref{fig:FSCDW}(a)), we find that the critical smectic field has a 
dynamic critical
exponent $z=2$, which will result in a $C\sim T$ contribution
to the heat capacity. This is a correction to the conventional linear $T$ behavior of Fermi liquids.  These quantum fluctuations lead to the existence of four points on the FS where the assumptions of FL theory are violated
(Fig. \ref{fig:FSCDW}(a)). At these points the imaginary part of the fermion self-energy correction 
$\Sigma^{\prime \prime}(k_F,\omega)\sim|\omega|^{1/2}$. For $Q_S=2k_F$  (Fig. \ref{fig:FSCDW}(b)), 
the system exhibits anisotropic scaling: $[q_x]=1$, $[q_y]=2$ and $[\omega]=3$
for the incommensurate CDW, while $[q_x]=1$, $[q_y]=2$ and $[\omega]=2$ for the commensurate case.
Besides, a non-analytic $\Phi^{5/2}$ term, where $\Phi$ is the smectic order parameter, is generated in the action of the low-energy effective theory. This non-analytic term is relevant under the renormalization group (RG) for the incommensurate case, suggesting a weak, fluctuation-induced, first-order transition.
This coupling is irrelevant in the commensurate case. Here we also find two points on the FS 
(Fig. \ref{fig:FSCDW}(b)), where the system has marginal
FL behavior, with a quasiparticle scattering rate $\Sigma^{\prime \prime} (k_F,\omega) \sim |\omega|$, and
a  low temperature correction to the heat capacity $C\sim T^{3/2}$, which is subleading. We also consider the special case of a CDW caused by a nearly nested FS, for which we find that the low-temperature heat capacity correction  $C\sim T^{4/3}$, which is also subleading, and the fermions form a  FL, with $\Sigma^{\prime \prime}(k_F,\omega) \sim |\omega|^{13/12}$. We also calculated the dynamic CDW susceptibility $\chi_S(q,\omega)$ for both cases.
The $Q_S>2k_F$ case will not be discussed here. In the presence of a lattice this case is quite trivial (see Sec. \ref{sec:N_S_critical}) while for it to occur in a continuum system, where it is non-trivial, requires unphysical assumptions.

The smectic phase is discussed in Sec. \ref{sec:smectic_phase}. In the smectic 
phase the anisotropic scaling associated with the Goldstone fluctuations are 
$[q_x]=1$, $[q_y]=[\omega]=2$. We find  that the low-temperature heat capacity correction $C\sim T^{3/2}$, which is also subleading. The quasiparticle scattering rate in this case is $\Sigma^{\prime \prime}(k_F,\omega) \sim \omega^2\log |\omega|$ for much of the FS while $\Sigma^{\prime \prime}(k_F,\omega) \sim |\omega|^{3/2}$ at the two special points where the Fermi velocity is parallel to the ordering wave vector. Thus, in this case fermions behave as a FL. We also calculated both the longitudinal and transverse  dynamic CDW susceptibilities in the smectic phase.

Lattice effects are also discussed. For the case of an incommensurate smectic 
phase, we show that there is an unpinned smectic phase close to the nematic-smectic
critical point. In this phase, the smectic Goldstone mode has a dynamic critical
exponent $z=1$ and the system is a FL with $\Sigma^{\prime \prime}(\omega) \sim \omega^2\log |\omega|$ at most of the FS and $\Sigma^{\prime \prime}(\omega)\sim \omega^2$ at some special point on the FS described below. 
Due to the unpinned smectic ordering, the system receives a correction to the low-temperature heat capacity $C \sim T^2$, and we also computed the dynamic transverse CDW susceptibility. Deep into the smectic
phase, an incommensurate CDW may be pinned down by lattice distortion. As expected, the fermions in a pinned smectic are in a conventional FL state.

In Sec. \ref{sec:finiteT} we present a brief discussion of the role of thermal fluctuations for these phases and of the classical-to-quantum crossovers. We conclude with a summary of our main results and a discussion of open questions in Sec. \ref{sec:conclusions}.
Details of the calculations are presented in several appendices. In Appendix \ref{app:sec:tensor} we discuss the tensor structure of the order parameters. In Appendix \ref{app:sec:critical} we present details of the nematic-smectic QCP for the case $Q_S<2k_F$, while the nonanalytic terms induced for the $Q_S=2k_F$ case are presented in Appendix \ref{app:sec:phi_nonanalytic}. In Appendix \ref{app:sec:gs_mode_smectic} we present details of the calculation of the spectrum of Goldstone modes in the smectic phase. In Appendix \ref{app:sec:RPA} we summarize the random phase approximation (RPA) calculation of  the fermion self-energy at the nematic-smectic QCP and in the smectic phase.

\section{Experimental Status of Electronic Liquid Crystal Phases}
\label{sec:expt}

During the past decade or so experimental evidence has been mounting of the existence of electronic liquid crystal phases in a variety of strongly correlated (as well as not as strongly correlated) electronic systems. We will be particularly interested in the experiments in the copper oxide {\hts}, in the  ruthenate materials (notably {\doubleRu}), and in two-dimensional electron gases (2DEG) in large magnetic fields. However, as we will discuss below, our results are also relevant to more conventional CDW systems such as the quasi-two-dimensional dichalcogenides. 

\subsection{{\HTS}}
\label{sec:hts}

In addition to {\hty}, the copper oxide materials display a strong tendency to have charge-ordered states, such as  stripes. The relation between charge ordered states\cite{kivelson-fradkin-2007}, as well as other proposed ordered states\cite{chakravarty01c,Varma2005}, and the mechanism(s) of {\hty} is a subject of intense current research. It is not, however, the focus of this paper. 

Stripe phases have been extensively investigated in {\hts} and detailed and recent reviews are available on this subject\cite{Kivelson2003,tranquada-2007}. Stripe phases in {\hts} have unidirectional order in both spin and charge (although not always)  and it is typically incommensurate. In general the detected stripe order (by low-energy inelastic neutron scattering) in {\LSCO}, {\LBCO} and {\YBCO} (see Refs.\cite{Kivelson2003} and \cite{tranquada-2007} and references therein) is not static but ``fluctuating''. As emphasized in Ref.\cite{Kivelson2003}, ``fluctuating order'' means that there is no true long range unidirectional order. Instead, the system is in a (quantum) disordered phase, very close to a quantum phase transition to such an ordered phase, with very low-energy fluctuations that reveal the character of the proximate ordered state. On the other hand, in {\LBCO} near $x=1/8$ (and in {\LNSCO} also near $x=1/8$),  the order detected  by elastic neutron scattering\cite{tranquada-2004}, and resonant x-ray scattering in {\LBCO} \cite{abbamonte-2005} also near $x=1/8$, becomes true long-range static order.

In the case of {\LSCO}, away from $x=1/8$, and particularly on the more underdoped side,  the in-plane resistivity has a considerable temperature-dependent anisotropy\cite{ando-2002}, which has been interpreted as an indication of electronic nematic order. From these experiments it has been suggested that this phase be identified as an electron nematic\cite{ando-2002}. The same series of experiments also showed that very underdoped {\YBCO} is an electron nematic as well. The most striking evidence for electronic nematic order in {\hts} are the recent  neutron scattering experiments in {\YBCO} at $y=6.45$\cite{hinkov-2007b}. In particular, the temperature-dependent anisotropy of the inelastic neutron scattering in {\YBCO} shows that there is a critical temperature for nematic order (with $T_c \sim 150 K$) where the inelastic neutron peaks also become incommensurate. Similar effects were reported by the same group\cite{hinkov-2006} at higher doping levels ($y\sim 6.6$) who observed that the nematic signal was decreasing in strength suggesting the existence of a nematic-isotropic quantum phase transition closer to optimal doping. Fluctuating stripe order in underdoped {\YBCO} has been detected earlier on in inelastic neutron scattering experiments \cite{mook-2000,stock-2004} which, in hindsight, can be reinterpreted as evidence for nematic order.   However, as doping increases the strength of the temperature-independent anisotropic background, due to the increased orthorhombicity of the crystal, also increases thus making this phase transition difficult to observe.

Recent inelastic neutron scattering experiments have found  similar effects in {\LSCO} materials where fluctuating stripes where in fact first discovered\cite{tranquada-1995}. Matsuda {\it et al} \cite{matsuda-2008} have given qualitatively similar evidence for nematic order in underdoped {\LSCO} ($x=0.05$) which was known to have ``fluctuating diagonal stripes''. In the same doping range it has also been found by resonant x-ray scattering experiments that 5\% Zn doping stabilizes a static diagonal stripe-ordered state with a very long persistence length which sets in at quite high temperatures\cite{abbamonte-2007}.

These recent results strongly suggest that the experiments that had previously identified the {\hts} as having  ``fluctuating stripe order'' (both inside and outside  the superconducting phase) were most likely detecting an electronic nematic phase, quite close to a state with long-range stripe (smectic) order. In all cases the background anisotropy (due to the orthorhombic distortion of the crystal structure) acts as a symmetry breaking field that couples linearly to the nematic order, thus rounding the putative thermodynamic transition to a state with spontaneously broken point group symmetry. These effects are much more apparent at  low doping where the crystal orthorhombicity is significantly weaker. 

The nature of the fluctuating spin order changes substantially as a function of doping: in the very underdoped systems there is no spin gap while inside much of the superconducting dome there is a finite spin gap. In fact in {\LBCO} at $x=1/8$ there is strong evidence for a complex stripe-ordered state which combines charge, spin and superconducting order\cite{li-2007,berg-2007}.
These experiments have also established that static long-range stripe charge and spin orders do not have the same critical temperature, with  static charge order having a higher $T_c$.

An important caveat to our analysis  is that in doped systems there is always quenched disorder, and has different degrees of short range ``organization'' in different {\hts}. Since disorder also couples linearly to the charge order parameters it ultimately also rounds the transitions and renders the system to a glassy state (as noted in Refs.\cite{Kivelson1998,Kivelson2003}). Such effects are evident in scanning tunneling microscopy (STM) experiments in {\BSCCO} which revealed that the high-energy (local) behavior of the {\hts} has charge order and it is glassy\cite{howald-2003a,Kivelson2003,hanaguri-2004,kohsaka-2007,vershinin04}.

Finally, we note that in the recently discovered iron pnictides based family of {\hts}, such as La (O$_{1-x}$F$_x$)FeAs \cite{kamihara-2008,mu-2008}, a unidirectional spin-density wave has been found. It has been suggested\cite{fang-2008} that the undoped system LaOFeAs may have a high-temperature nematic phase and that quantum phase transitions also occur as a function of fluorine doping\cite{xu-2008}. This suggests that many of the ideas and results that we present here may be relevant to these still poorly understood materials.

\subsection{Other complex oxides}
\label{sec:other}

The existence of stripe-ordered phases is well established in other complex oxide materials, particularly the manganites and the nickelates. In general, these materials tend to be ``less quantum  mechanical'' than the cuprates in that they are typically insulating (although with interesting magnetic properties) and the observed charge-ordered phases are very robust. These materials typically have larger electron-phonon interactions and electronic correlations are comparatively less dominant in their physics. For this reason they tend to be ``more classical'' and less prone to quantum phase transitions. However, at least at the classical level, many of the issues we discussed above, such as the role of phase separation and Coulomb interactions, also play a key role\cite{dagotto-2001}. The thermal melting of a stripe state to a nematic has been seen in the manganite material Bi$_x$Ca$_x$MnO$_3$\cite{rubhausen-2000}.

\subsection{Ruthenates}
\label{sec:ruthenates}

Recent magneto-transport experiments in the quasi-two-dimensional bilayer ruthenate {\doubleRu} by the St. Andrews group\cite{Borzi2007} have given strong evidence of a strong temperature-dependent in-plane transport anisotropy in these materials at low temperatures  $T\lesssim 800$ mK and for a window of perpendicular magnetic fields around $7.5$ Tesla. These experiments provide strong evidence that the system is in an electronic nematic phase in that range of magnetic fields\cite{Borzi2007,fradkin-2007}. The electronic nematic phase appears to have preempted a metamagnetic QCP in the same range of magnetic fields\cite{Grigera01,millis-2002,Perry04,Green05}. This suggests that proximity to phase separation may be a possible microscopic mechanism to trigger such quantum phase transitions, consistent with recent ideas on the role of Coulomb-frustrated phase separation in 2EDGs\cite{jamei-2005,lorenzana-2002b}.

\subsection{2DEGs in large magnetic fields}
\label{sec:2DEG}

To this date, the best documented electron nematic state is the anisotropic compressible state observed in 2DEGs in large magnetic fields near the middle of a Landau level, with Landau index $N \geq 2$\cite{Lilly1999,lilly-1999b,Du99,Pan99}. In ultrahigh-mobility samples of a 2DEG in AlAs-GaAs heterostructures,  transport experiments in the second Landau level (and above) near the center of the Landau level  show a pronounced anisotropy of the longitudinal resistance rising sharply below $T \simeq 80$ mK, with an anisotropy that increases by orders of magnitude as the temperature is lowered. These experiments were originally interpreted as evidence for a quantum Hall smectic (stripe) phase \cite{koulakov96,moessner96,Fradkin99,macdonald-fisher-2000,barci-2002a}. Further experiments\cite{Cooper01,cooper-2002,Cooper03}
did not show any evidence of pinning of this putative unidirectional CDW as the $I-V$ curves are strictly linear at low bias and no broadband noise was detected. In contrast, extremely sharp threshold electric fields and broadband noise in transport was observed in a nearby reentrant integer quantum Hall phase, suggesting a crystallized electronic state. These facts, together with a detailed analysis of the experimental data, suggested that the compressible state is in an electron nematic phase\cite{Fradkin99,Fradkin00,wexler-2001,Radzihovsky02,doan-manousakis-2007}, which is better understood as a quantum melted stripe phase.

\subsection{Conventional CDW materials}
\label{sec:cdw}

CDWs have been extensively studied since the mid-seventies and there are extensive reviews on their properties\cite{Gruner1988,Gruner1994}. From the symmetry point of view there is no difference between a CDW and a stripe (or electron smectic). The CDW states are usually observed in systems which are not particularly strongly correlated, such as the quasi-one-dimensional and quasi-two-dimensional dichalcogenides, and the more recently studied tritellurides. These CDW states are reasonably well described as FLs which undergo a CDW transition, commensurate or incommensurate,  triggered by a nesting condition of the FS\cite{mcmillan75,McMillan1976}. As a result, a part or all of the FS is gapped in which case the CDW may or may not retain metallic properties. Instead, in a strongly correlated stripe state, which has the same symmetry breaking pattern, at high energy has Luttinger liquid behavior\cite{Kivelson1998,Emery00,carlson04}.

What will interest us here is that conventional quasi-2D dichalcogenides, the also quasi-2D tritellurides and other similar CDW systems can quantum melt as a function of pressure in {\TiSe2}\cite{snow03}, or by chemical intercalation as in Cu$_x${\TiSe2}\cite{morosan-2006,Barath2007} and Nb$_x$TaS$_2$\cite{lieber-1991}. Thus, CDW phases in chalcogenides can serve as a weak-coupling version of the problem of quantum melting of a quantum smectic. Interestingly, there is strong experimental evidence that both {\TiSe2}\cite{snow03} and Nb$_x$TaS$_2$\cite{lieber-1991} do not melt directly to an isotropic Fermi fluid but go instead through an intermediate phase, possibly hexatic. (Cu$_x${\TiSe2} is known to become superconducting\cite{morosan-2006}.) Whether or not the intermediate phases are anisotropic is not known as no transport data is available in the relevant regime. 

The case of the CDWs in tritellurides is more directly relevant to the theory we present in this paper. Tritellurides are quasi-2D materials which for a broad range of temperatures exhibit a unidirectional CDW ({\it i.e.\/} an electronic smectic phase) and whose anisotropic behavior appears to be primarily of electronic origin\cite{brouet-2004,laverock-2005,sacchetti-2006,sacchetti-2007,fang-2007}. However, the quantum melting of this phase has not been observed yet. Theoretical studies have also suggested that it may be possible to have a quantum phase transition to a state with more than one CDW in these materials\cite{yao-2006}.

\section{Order-Parameter Theory}
\label{sec:classical}

In this section we will construct, using phenomenological arguments, an effective order parameter theory that will describe both the electron nematic and the electron smectic (or unidirectional CDW) phases. Although by symmetry the order parameter theory must be very similar to the ones used in classical
liquid crystal phases, we will go through the construction of 
the phenomenological theory in some detail for several reasons.
In 2D the rotation group $SO(2)$ is Abelian which allows for
a significant simplification of the formulas by using a complex order parameter 
for the nematic phase, instead of a tensor expressions commonly used for 3D
classical liquid crystals. Proper dynamical terms now need to be included 
to describe the quantum fluctuations at zero-temperature.
Besides, in order to provide a clear relation between this paper and earlier 
studies of the CDW state of fermions, we would like to discuss also the 
relation between the smectic phase and the CDW state.

\subsection{The normal-electronic nematic transition}
\label{sec:normal-nematic}

The nematic order parameter in 2D is a $l=2$ representation of the $SO(2)$ rotational
group \cite{Oganesyan2001}. It is defined as a symmetric traceless tensor of rank two.
\begin{align}
\mathbf{N}=\left(
\begin{array}{ccc}
n_{11} & n_{12}
\\
n_{12} & -n_{11}
\end{array}
\right).
\label{eq:2D_nematic}
\end{align}
The 2D rotational group $SO(2)$ is isomorphic to $U(1)$. Hence, we
define instead the complex order-parameter field $N(\vec{r},t)$
\begin{align}
	N(\vec{r},t)=n_{11}(\vec{r},t)+i n_{12}(\vec{r},t),
\label{eq:2D_nematic_complex}
\end{align}
where $\vec{r}$ and $t$ are the space and time coordinates. We will use 
this complex order parameter field in this paper to take the advantage 
of the Abelian nature of $SO(2)$. 

The conjugate field is
$N^\dagger(\vec{r},t)=n_{11}(\vec{r},t)-i n_{12}(\vec{r},t)$.
Under a global rotation by an angle $\theta$, the fields $N(\vec r, t)$ and $N^\dagger(\vec r, t)$  transform, respectively, as $N(\vec r, t) \to e^{2 i \theta}\; N(\vec r, t)$ and 
$N^\dagger(\vec r, t) \to e^{-2 i \theta} N^\dagger(\vec r, t)$. Hence,
$N$ and $N^\dagger$ carry the angular momentum quantum numbers $l_z=2$ and $l_z=-2$, respectively.

This complex order parameter can be generalized easily to other angular 
momentum channels $l\ne2$, but not to higher dimensions $d>2$, since it 
relies heavily on the special property of the 2D rotational group $SO(2)$. 
In higher dimensions, the rotational group will no longer be Abelian, so 
one will need to use the tensor formula as in the classical liquid crystal theories. 
In Appendix \ref{app:sec:tensor}, formulas using the complex order 
parameter are translated into the conventional tensor form for comparison.

The order-parameter field we just defined is invariant under spatial-inversion 
and  time-reversal
\begin{align}
P N(\vec{r},t) P^{-1}=N(-\vec{r},t),
\\
T N(\vec{r},t) T^{-1}=N(\vec{r},-t).
\end{align}
In even spatial dimensions, including 2D in which our system lives, a 
chiral transformation is different from a space inversion. To change the 
chirality in 2D, we can reverse the $y$ direction and keep the $x$ 
direction unchanged. Under this chiral transformation, the nematic field 
will be changed into the conjugate field
\begin{align}
C N(x,y,t) C^{-1}=N^{\dagger}(x,-y,t).
\end{align}
Here, $C$ is the chiral transformation operator. 

The effective action must preserve the symmetries of the system, both continuous, as the 
translational and rotational symmetries, and discrete, as the time reversal,
spatial inversion and chiral symmetries. With the assumption of analyticity, 
the action must be
\begin{align}
S_N&=&\int \mathrm{d}\vec{r}\mathrm{d}t 
\left(|\partial_t N|^2-|\vec{\nabla} N|^2-\Delta_N |N|^2
-u_N |N|^4\right).
\nn\\&&
\label{eq:N_action}
\end{align}
Here the dynamical term is quadratic in time derivatives. This is because the term
linear in time derivatives $-i N^\dagger \partial_t N+ h.c.$
is not allowed by the chiral symmetry. It is the imaginary part of 
$N^\dagger \partial_t N$. It corresponds to a pseudoscalar, and is not allowed.

In 2D, cubic terms in the nematic field $N$ are not allowed. Hence, if $u_N>0$, the normal-nematic 
transition is second order, instead of a first-order transition as in the 3D case \cite{DeGennes1993,Chaikin98}.
For $u_N>0$ and $\Delta_N>0$, the rotational invariant ground state will be stable.
When $\Delta_N$ becomes negative, $N$ will develop an expectation value $\bar{N}$
with module $\sqrt{-\Delta_N/(2 u_N)}$, which breaks the $SO(2)$ rotation symmetry. 
The residual rotational symmetry would be $Z_2$. The argument of $\bar{N}$ 
determines the direction of the nematic order parameter.

The action of Eq. \eqref{eq:N_action} has an internal $U(1)$ symmetry 
associated with the phase of the complex field $N$, which is not physical. 
By symmetry, terms of the form
\begin{align}
-\kappa \left( N^\dagger (\partial_x+i \partial_y) 
N^\dagger (\partial_x+i \partial_y) N+h.c.\right)
\label{eq:frank_term_complex}
\end{align}
are allowed \cite{Oganesyan2001,chiral}. This kind of terms are irrelevant at the 
QCP and in the isotropic phase, which leads to the  existence of an ``emergent'' internal  $U(1)$ symmetry at quantum criticality.  
But it will be important in the nematic phase, as it makes the two Frank 
constants to attain different values. (This effect is formally analogous to the role of spin-orbit interactions in the Schr\"odinger equation: in their absence spin is an internal degree of freedom.)
This emergent symmetry of the normal phase and at the critical point is very important 
for the classical normal-nematic transition, especially in the study about the 
fluctuation effects \cite{Priest1976,Korzhenevskii1979, Nelson1981}.

\subsection{The electronic nematic phase}
\label{sec:nematic}

In the nematic phase, the $SO(2)$ rotational symmetry is broken.
Hence, we expect the fluctuations of the amplitude of the nematic order parameter, $\delta N$, to correspond to a massive 
mode with an energy gap of $-2 \Delta_N$ ($\Delta_N<0$), and the fluctuations of 
the phase, $\phi_N$, constitute the gapless Goldstone mode. Without loss of generality, 
throughout this paper, we assume that $\bar N$ is real and positive. This state 
corresponds to a nematic order in the main axis direction. In this state, the action 
of $\phi_N$ is
\begin{eqnarray}
&&\!\!\!\!\!\!\!\!\! S_{\phi_N}\!=\!\bar{N}^2\!\int d\vec{r}dt\;
\left((\partial_t\phi_N)^2\!-\!K_1 (\partial_x \phi_N)^2-K_2 (\partial_y \phi_N)^2\right).\nonumber \\
&&
\label{eq:nematic_goldstone}
\end{eqnarray}
where $K_{1}=1 + 2 \kappa \bar N$ and $K_2=1-2\kappa \bar N$ are the two Frank constants.
This action is only valid for small nematic fluctuations. It cannot be 
used to study topological defects of the nematic phase, known as 
disclinations\cite{DeGennes1993}.
The field $\phi_N$ has dynamic critical exponent $z=1$. This makes the 
effective dimension of this system $3$, which is above the lower critical 
dimension of the theory $d=2$, and nematic order will
not be destroyed by fluctuations.

\subsection{CDW multi-critical point}
\label{sub:sec:cdw-multi}

The smectic order is a unidirectional CDW, described by a single complex order parameter field. If we assume analyticity, the 
effective low-energy theory of the bosonic field $\rho$ can be determined as:
\begin{align}
S_{\textrm{CDW}}=\int \mathrm{d}\vec{r}\mathrm{d}t \; (\partial_t \rho) ^2
+S_{2}+S_3+S_4,
\label{eq:CDW_action}
\end{align}
where $S_2$ is the term in the quadratic order of $\rho$, $S_3$ and $S_4$
are the cubic and quartic terms respectively.

The term $S_{2}$ in the momentum space is,
\begin{align}
S_{2}=-\int\frac{\mathrm{d}\vec{k}\mathrm{d}\omega}{(2\pi)^3} 
f(|k|) \rho(\vec{k},\omega) \rho(-\vec{k},-\omega).
\label{eq:CWD_action_quadratic}
\end{align}
The function $f(|k|)$ has the physical meaning of the inverse of the 
CDW susceptibility. If we assume the ordering wave-vector of the CDW
is $\vec Q_S$ (with $|\vec Q_S|=Q_S$ its magnitude), $f(|\vec k|)$ will have the form
\begin{equation}
f(|\vec k|)=\Delta_{\textrm{CDW}}+C (|\vec k|^2-Q_S^2)^2 +\ldots
\label{eq:inverse_CDW_susceptibility}
\end{equation}
where $\Delta_{\textrm{CDW}}$ is the energy gap of the CDW excitations
\cite{Brazovskii1975}, and $C$ is a positive constant.

When the energy gap $\Delta_{\textrm{CDW}}$ decreases 
to zero, all the density wave modes with $|\vec k|=Q_S$ will become soft and 
critical when $\Delta_{\textrm{CDW}}=0$.
This is very different from an ordinary $\phi^3$- or $\phi^4$-theory, where we
only need to consider one mode (or two modes for a complex field) at small momentum.
Here, we need to consider all the modes with the wave vector $\vec k$ whose magnitude is close to 
$Q_{S}$. In other words, the point
$\Delta_{\textrm{CDW}}=0$ is not a critical point but a multi-critical point 
with an infinite number of critical modes. Even if a lattice background is 
present, the $\Delta_{\textrm{CDW}}=0$ point may still be a multi-critical 
point of $n$ critical modes, if the lattice has a $n$-fold rotational symmetry 
for $n>2$. 
For a multi-critical point, higher-order terms become 
important. Without a detailed knowledge of these higher-order terms, it is not
possible to determine whether the transition is first or second order, or how many 
CDWs will be formed in the ordered phase.

Brazovskii \cite{Brazovskii1975} studied the classical version of this problem, 
considering only the isotropic interactions.
Chubukov and co-workers \cite{Chubukov2005} studied the quantum problem
in a fermionic system in the high-density regime where the cubic and quartic 
terms of $\rho$ can be ignored.

In general, depending on the non-Gaussian terms, the ordered phase may have only 
one or several CDWs \cite{Brazovskii1975}. For a rotational invariant 
system, it is often assumed that $3$ CDWs form a triangular lattice to 
minimize the breaking of the rotational symmetry, as the 2D Wigner crystal 
state \cite{Wigner1934}. For systems with a strong lattice potential, the 
system is often assumed to become an electron crystal state which preserves the point 
group rotational symmetry of the background lattice, {\it e.g.\/} the rare-earth tritellurides\cite{yao-2006}.

\begin{figure}[h!]
\begin{center}
\includegraphics[width=0.45\textwidth]{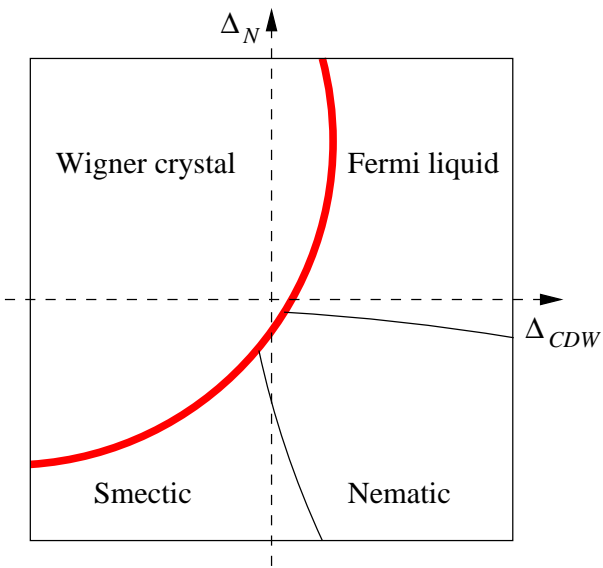}
\end{center}
\caption{(online color) Schematic phase diagram at $T=0$ as a function of $\Delta_{N}$ and
$\Delta_{CDW}$ defined in Eqs.\eqref{eq:N_action} and \eqref{eq:inverse_CDW_susceptibility}. The cross point of the two dash lines is the multi-critical point
$\Delta_{N}=\Delta_{CDW}=0$. The red thick lines stand for first order phase
boundaries. Other phase boundaries may be first or second order. More complex 
electron crystal phases are possible, for example, an anisotropic
electron crystal phase where more than one CDWs and nematic coexist, but they are 
beyond the discussion of this paper.}\label{fig:phase_diagram}
\end{figure}

For isotropic systems, outside the nematic phase, the $\rho^3$ term in Eq. \eqref{eq:CDW_action} favors that
three CDWs form by a first-order transition \cite{Brazovskii1975}. However, inside a nematic phase, as we will show below, the nematic order parameter, which is coupled to $\rho^2$, 
favors only one CDW and will compete with the $\rho^3$ term. For a continuous quantum phase
transition, $\rho^3$ will be a subleading perturbation
compared to $\rho^2$, at least close enough to the transition. Hence, the smectic phase, a unidirectional CDW,  will be energetically favorable. On the other hand, in the case of a first-order transition, depending on  
microscopic details either the 
smectic phase or the state with three CDWs would be preferred. We represent these different possibilities in the schematic phase diagram shown in Fig.
\ref{fig:phase_diagram}.

On a square lattice, due to the point group symmetry of the lattice, the electron crystal phase  usually consists of $2$ CDWs perpendicular to each
other. The phase transition between this phase and the FL may be second order due to the absence of the cubic term $\rho^3$ which, in contrast to isotropic systems, is prohibited by momentum conservation. We have confirmed this structure of the phase diagram in a microscopic mean-field calculation. However, at the multi-critical point where both the CDW modes and the nematic mode are critical (the $(0,0)$ point in Fig. \ref{fig:phase_diagram}), the coupling between CDWs and the nematic order parameter (Eq. \eqref{eq:s_n_coupling_complex}) is relevant. This suggests a fluctuation driven first-order transition near the multi-critical point. Hence, this multi-critical point is essentially unreachable.

In this paper, we study the nematic-smectic phase transition and the smectic phase 
using a weak coupling approach by perturbing about a FL state. This approach is consistent
provided the nematic phase is narrow enough in coupling constant space so that the nematic-smectic transition is not too
far from the FL phase.

It is useful to compare to the classical version of this problem. The theory of classical (thermal) melting in two dimensions, the Kosterlitz-Thouless-Halperin-Nelson-Young theory\cite{nelson-1979,young-1979} (see Ref.\cite{Chaikin98}), is a theory of a phase transition driven by the proliferation of topological defects: a dislocation unbinding transition  in the case of melting of a 2D Wigner crystal (a triangular lattice)  into a hexatic phase, and  disclination unbinding transition in the hexatic-isotropic phase transition. (The case of the square lattice was discussed only recently in Ref.\cite{delmaestro-2005}).  The reason for the success of the classical theory of melting in two dimensions is that, as in all Kosterlitz-Thouless phase transitions\cite{kosterlitz-1973,Chaikin98}, at finite temperatures the classical ordered state with a spontaneously broken continuous symmetry is not possible in two dimensions. Instead, there is a line (or region) of classical critical behavior with exactly marginal operators. The defect-unbinding phase transition appears as an  irrelevant operator  becoming marginally relevant. 
  
 In the case of the quantum phase transitions in two dimensions that we are interested in, there are no such exact marginal operators available  at zero temperature, and hence, no lines of fixed points available. Thus, the $T=0$ quantum phase transition is not triggered by a defect-unbinding operator becoming marginal, but instead by making the coupling constant of an irrelevant operator large (as in standard continuous phase transitions, classical or quantum). Instead, the quantum phase transition is closer to Landau-type (or, rather, Hertz-Millis like) description in that it is governed (as we will see) by a quantum-mechanical analog of the celebrated  McMillan-deGennes theory for a nematic-smectic phase transition in classical liquid crystals in three dimensions\cite{DeGennes1993,Chaikin98}. The approach that we will pursue here does not contain much of the physics of strong correlations as it begins with a state with well-defined fermionic quasiparticles. It also does not treat correctly the tendency of strongly correlated systems to exhibit inhomogeneous states and phase separation. The only way to account for this physics correctly is to use the opposite approach, a strong coupling theory of quantum melting of the crystal and stripe phases, as advocated in Ref.\cite{Kivelson1998}. So far, this theory only treats the physics deep inside a stripe phases, and the theory of their quantum melting to a nematic phase does not yet exist.
 Thus, although from a strong-coupling perspective it would be highly desirable to have such a defect unbinding theory of this quantum phase transition (such a description does exist for an insulating system\cite{zaanen-2004} but its extension to a metallic state is not available and it is highly non-trivial), we will pursue instead a Hertz-Millis approach \cite{Hertz1976,Millis1993,Sachdev1999} to this quantum phase transition.

\subsection{The electronic nematic-smectic transition}
\label{sec:nematic-smectic}

Nematic order will remove the degeneracy of CDW modes in different 
directions and select one CDW. As a result, the Brazovskii CDW multi-critical point becomes just a critical point.
For simplicity, we assume that the nematic order parameter is small enough so that a Landau-type expansion still makes sense, which is equivalent to assuming that the system is still ``close enough'' to the nematic-isotropic QCP. However, as we will show later, the critical theory we get using these assumptions has the only form allowed by symmetry, assuming analyticity.

By symmetry, the coupling between the CDW and the nematic field 
is
\begin{widetext}
\begin{align}
S_{\textrm{int}}= -g \int \frac{\mathrm{d}\vec{k}\mathrm{d}\Omega}{(2\pi)^3}
\int \frac{\mathrm{d}\vec{q}\mathrm{d}\omega}{(2\pi)^3}
N (\vec{q},\omega) e^{-2 i \theta_k}
\rho(\vec{k}-\vec{q},\Omega-\omega) 
\rho (-\vec{k},-\Omega)
+\textrm{h.c.}
\label{eq:s_n_coupling_complex}
\end{align}
\end{widetext}
whose tensor form is shown in Appendix \ref{app:sec:tensor}. Here, $\theta_k$ is the polar angle of $\vec{k}$.
This term is irrelevant in the isotropic phase, but in the nematic phase, 
where $N$ gets the expectation value $\bar N$; this term will be 
of the same order as $S_2$, which was defined in Eq. 
\eqref{eq:CWD_action_quadratic}, and hence  it becomes important.

Inside the nematic phase the amplitude fluctuations of the nematic order parameter are gapped while the orientational fluctuations, the nematic Goldstone modes, are gapless, at least strictly in the absence of a lattice and other orientational symmetry breaking couplings. Thus, deep enough in the nematic phase it is possible to integrate out the gapped nematic amplitude fluctuations and derive an effective theory involving the gapless nematic Goldstone mode. However, as the nematic-smectic phase transition is approached, the gap of the fluctuations of the smectic order parameter will get smaller and will approach zero at the QCP. Thus, in this regime, the nematic phase has low-energy ``fluctuating stripes''. This regime is the analog of that in conventional liquid crystals where the McMillan-deGennes classical theory applies . We will now see how this theory arises in the quantum case.

The leading term in $S_{\textrm{int}}$ of Eq. 
\eqref{eq:s_n_coupling_complex} will be
\begin{align}
-2 g \int \frac{\mathrm{d}\vec{k}\mathrm{d}\omega}{(2\pi)^3}
\bar N \cos(2 \theta_k)
\rho (\vec{k},\omega) \rho (-\vec{k},-\omega).
\end{align} 
This term will stabilize the density wave in either $x$ or $y$ direction and 
destabilize the other, depending on the sign of $g$. As a result, the nematic order 
will select a special direction along which only one CDW will form. Past this phase transition
the system will be in a smectic state, a unidirectional CDW. For simplicity, we assume $g>0$, which selects $\vec{Q}_S$ in the $y$ 
direction.

Only the density fluctuations close to $\vec{k}=\pm \vec{Q}_S$ matter for the low-energy theory. We define a complex field $\Phi$, 
describing the density fluctuations around $\vec{k}=\pm\vec{Q}_S$ as
\begin{align}
\Phi(\vec{q},\omega)=\rho(\vec{q}+\vec{Q}_S,\omega),
\end{align}
where $q$ is small. The real part of $\Phi(\vec{r},t)$ measures the density
fluctuations.

Under a spatial inversion, $\Phi(\vec{r},t)$ will become its conjugate field 
$\Phi^\dagger(-\vec{r},t)$. Hence, the term $-i\Phi^\dagger \partial_t \Phi+h.c.$
is not allowed in the Lagrangian, and the dynamical term 
for $\Phi$ is at least quadratic in time derivatives.

The cubic term of the field in the isotropic-CDW transition vanishes in the 
nematic-smectic transition, due to momentum conservation. By 
expanding Eq. \eqref{eq:s_n_coupling_complex} around $q\sim 0$, 
$\vec{k}\sim \vec{Q}_S$ and $\phi_N\sim0$, we obtain
\begin{align}
S=&S_{\phi_N}+\int \mathrm{d}\vec{r}\mathrm{d}t \Big( |\partial_t \Phi |^2
- C_y |\partial_y \Phi|^2
\nn\\
&
 -C_x |(\partial_x -i \frac{Q_S}{2} \phi_N)\Phi|^2
- \Delta_S |\Phi|^2 
-u_{S} |\Phi|^4\Big).
\nn\\
\label{eq:ns_action}
\end{align}
Here $S_{\phi_N}$ is the action of the nematic Goldstone mode defined in Eq. 
\eqref{eq:nematic_goldstone}.

The action of Eq. \eqref{eq:ns_action} is just a 2D version McMillan-de Gennes 
theory of the nematic-smectic transition in the classical liquid crystals but with $z=1$
quantum dynamics. The constants in  Eq.\eqref{eq:ns_action} are
\begin{flalign}
C_x&=\displaystyle{\frac{4 g \bar{N}}{Q_S^2}},&                          C_y&=C
\nonumber \\
\Delta_{S}&=\Delta_{\textrm{CDW}}-2 g \bar{N},& u_{S}&=u_{\textrm{CDW}}-\displaystyle{\frac{4 g^2}{\Delta_N}}.
\end{flalign}
Here $\Delta_{S}$ is the energy gap of $\Phi$ field, which mainly comes from the
CDW gap defined in Eq. \eqref{eq:inverse_CDW_susceptibility}. The correction term
$-2 g \bar{N}$ comes from the nematic ordering. The $u_{S}$ term comes from
the interactions between CDWs and it gets a correction from the amplitude
fluctuations of the nematic order, which has been integrated out. 
The nematic Goldstone field $\phi_N$ couples to the CDW field $\Phi$ as a gauge field with a
``charge'' $Q_S/2$. Here the two in the denominator comes from the fact that 
the nematic order parameter has an angular momentum $l=2$.
This gauge-like coupling is required by the rotational symmetry since, under spatial rotation by a small angle $\theta$, the fields transform as $\Phi\rightarrow \exp(i Q_S \; x\; \theta)\; \Phi$ and
$\phi_N\rightarrow \phi_N+l \theta$ (for the angular momentum channel $l$).
In fact, with the symmetry constrain and the assumption of analyticity, 
the action we show in Eq. \eqref{eq:ns_action} is the only allowed form for the effective low-energy theory, provided the topological excitations of $\phi_N$ are ignored 
\cite{Renn1988}.
Therefore, although we only keep linear terms of $\bar N$ in our calculations above, 
which is valid close to the normal-nematic critical point, the action in Eq. 
\eqref{eq:ns_action} will have the same form even deep inside the nematic phase.

The theory with the effective action given in Eq.\eqref{eq:ns_action} has a critical field $\Phi$ and gapless Goldstone boson $\phi_N$. A naive mean-field theory would suggest that this is a continuous phase transition. In the case of the theory of classical liquid crystals, where the same naive argument also holds,  Halperin, Lubensky and Ma
\cite{Halperin1974} used the $4-\epsilon$ expansion to show that there is a run-away behavior in the renormalization group flows, 
similar to that of superconducting transition coupled to a fluctuating electromagnetic field. They concluded that in both  cases the transition is probably weakly first order, a fluctuation-induced first-order transition. In other terms, in the classical theory the coupling of the smectic to the nematic Goldstone mode (which has the same form as a coupling to a gauge field) is relevant. To ascertain what happens in the case of the metallic nematic-smectic QCP  we will also need to take into account the effects of the fermionic degrees of freedom. We will see that  the fermionic fluctuations change the critical behavior in an essential way.

\subsection{The electronic smectic phase: a unidirectional CDW}
\label{sec:smectic}

In the smectic phase, the amplitude fluctuations of the order parameter,
$\delta \Phi$, are gapped but the phase fluctuations, $\phi_\Phi$, are
gapless, as required by the Ward-identity. This happens in systems for which lattice effects can be neglected, and hence are described formally in a continuum, or if the smectic order is sufficiently incommensurate. Therefore, upon integrating out the gapped amplitude fluctuations $\delta \Phi$,  the effective low-energy theory of the Goldstone mode 
becomes
\begin{align}
S_{\phi_\Phi}=\int \mathrm{d} \vec{r}\mathrm{d}t\;
	\left[\kappa_0(\partial_t \phi_\Phi)^2-\kappa_1(\partial^2_x \phi_\Phi)^2
	-\kappa_2 (\partial_y \phi_\Phi)^2
	\right],
\label{eq:class_smectic_goldstone}
\end{align}
When we are close to the nematic-smectic
critical point, the coefficients of this effective action are
\begin{align}
\kappa_0=|\bar{\Phi}|^2,
\qquad
\kappa_1=\displaystyle{\frac{4 K_1 \bar N^2}{Q_S^2}},
\qquad
\kappa_2=C_y |\bar{\Phi}|^2,
\label{eq:kappas}
\end{align}
with $\bar{\Phi}$ being the expectation value of the CDW order parameter. 
The vanishing of the stiffness 
$(\partial_x \phi_\Phi)^2$ term \cite{Peierls1935,Landau1937} is 
required by the Ward identity of rotational invariance. Thus, an underlying lattice, which will break the continuous rotational symmetry down to its discrete point group, will lead to a non-vanishing stiffness. Nevertheless, in many cases and particularly away from situations in which the FS is strongly nested, the breaking of rotational invariance can be parametrically small enough that at low temperatures its effects to a first approximation can be neglected and treated perturbatively afterward.

A simple scaling analysis of the effective action of Eq. \eqref{eq:class_smectic_goldstone} shows that, at the tree level, the scaling dimensions of space and time $[x]$, $[y]$ and $[t]$, are  $-1$, $-2$ and $-2$, respectively. Although the time direction and the
$y$ direction scales in the same way, the $x$ and $y$ directions now scale 
differently. This is a typical phenomenon for anisotropic states. Now the 
effective dimensions of this quantum theory is $5$. Hence,
our theory is above its (upper) critical dimension. So the higher-order 
interactions of the smectic Goldstone mode will be irrelevant, if we don't consider 
topological defects. The fact that we are above the critical dimension 
also tells us that the quantum fluctuations in the quantum smectic phase will 
not destroy the long-range order.

This scaling is very different from the classical smectic phase of 3D,
where $[x]=[y]=-1$ and $[z]=-2$, if the modulation is on the $z$ direction.
This classical theory is at its lower critical dimension, and long-range 
order is destroyed by fluctuations \cite{Peierls1935,Landau1937}, 
resulting in a power-law 
quasi-long-range order. This system has a line of critical points, so the higher order terms of the action that need to be considered 
were found to lead to logarithmic corrections to the power-law behavior 
\cite{Grinstein1981}.

The above analysis implies that our quantum problem is above the lower critical dimension. Therefore all 
these effects
of the 3D classical smectic phase will not be present in the 2D 
quantum case. The scaling behavior of a 2D quantum system
is similar to the columnar state of the classical liquid crystals,
instead of that of classical smectics. The classical
columnar state has two density waves so that it is a solid in two directions
but a liquid in the third direction. The Goldstone fluctuations of 
this state scale as $[x]=-1$ and $[y]=[z]=-2$ \cite{DeGennes1993},
which is the same as in the present case, if we consider the time direction 
in our problem as the $z$
direction. The difference between the classical columnar state  and the 2D quantum smectic state is that in the 3D 
columnar state, the Goldstone mode is a planar vector but in the present
case it is a scalar.

\section {Coupling the Order Parameter Theory to Fermions}
\label{sec:effect_fermions}

\begin{figure}
\begin{center}
\subfigure[~$Q_S<2k_F$]{\includegraphics[width=0.22\textwidth]{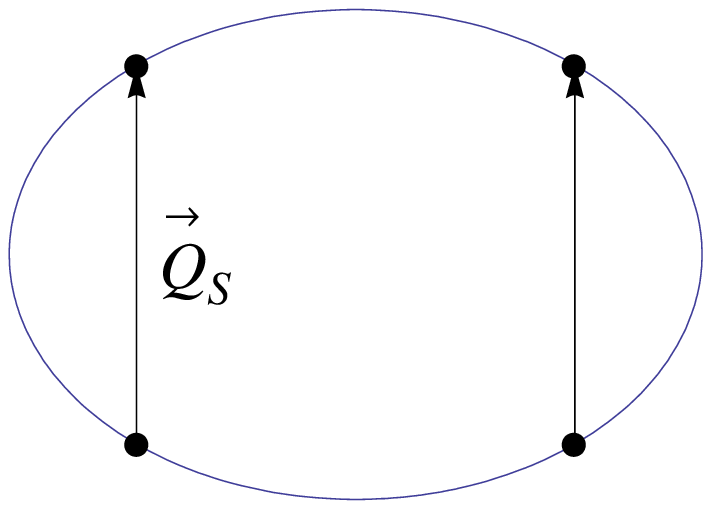}}
\subfigure[~$Q_S=2k_F$]{\includegraphics[width=0.22\textwidth]{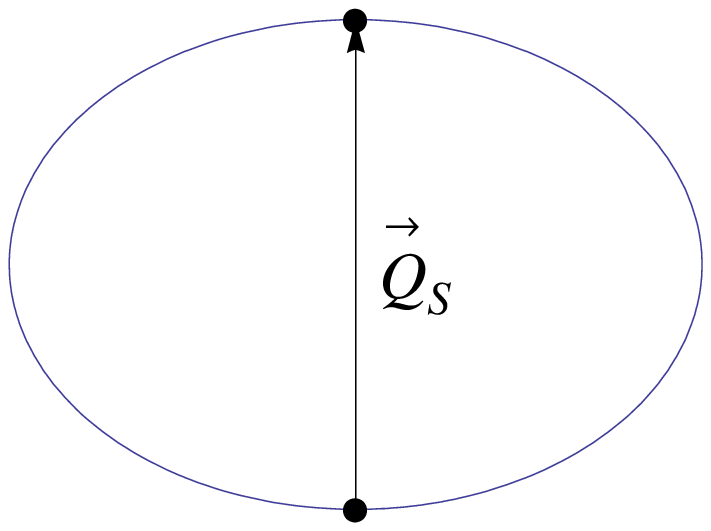}}
\subfigure[~$Q_S<2k_F$]{\includegraphics[width=0.22\textwidth]{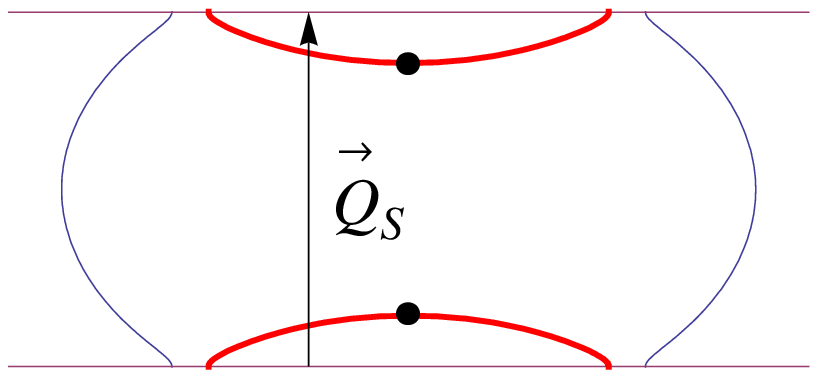}}
\subfigure[~$Q_S=2k_F$]{\includegraphics[width=0.22\textwidth]{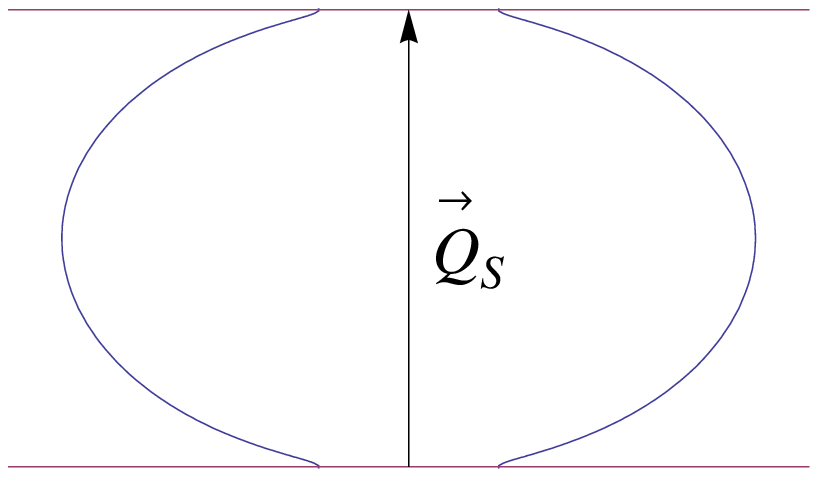}}
\end{center}
\caption{(online color) The FS of the nematic phase (a and b) and the reconstructed FS in the smectic phase (c and
d). (a) and (c) are for $Q_S<2k_F$ at the QCP and in the smectic phase respectively, while (b) and (d) are $Q_S=2k_F$, also at the QCP and in the smectic phase respectively. In (a) and (b), the black dots marked the non-FL points on the FS caused by the smectic mode fluctuations at the nematic-smectic QCP. The relevance of the points in (c) is explained in Sec. \ref{sec:smectic_phase}. In (c) we have show the case of $Q_S$ to be comparable to $2kF$ so as to keep the FS reconstruction simple. Here we show the effective Brillouin Zone with an open orbit and a closed pocket. The reconstructed FS of case (d) is partially gapped and the FS has an open orbit.
}\label{fig:FSCDW}
\end{figure} 

We will now proceed to couple the phenomenological theory of the nematic and smectic phases to a system of {\it a priori} well-defined fermionic quasiparticles described by the Landau theory of the FL. In a fermionic liquid crystal state, the bosonic order-parameter fields, defined above, 
will couple to the fermions. 

Let us define $\psi^\dagger(x,t)$ and $\psi(x,t)$ 
to be the fermion creation and annihilation 
operators of a FL. We will assume that the FL has a well-defined FS, which for simplicity we will assume is circular. (For lattice systems the FS will have the symmetry of the point group of the lattice.) The Fermi wave vector 
is $k_F$. The Fermi velocity is set to $1$ so that the energy and momentum  
have the same units. Consistent with the assumptions of the Landau theory of the FL\cite{Baym91} the effective Hamiltonian of the fermionic quasiparticles will be taken to be that of a free Fermi system, with a well-defined FS, and a set of quasiparticle interactions parametrized by the Landau parameters. These interactions are irrelevant in the low-energy limit of the FL but play an important role in the physics of electronic liquid crystal phases \cite{Oganesyan2001}. In any case in our discussion it will be unnecessary to include the Landau parameters explicitly since their effects will already be taken into account through the coupling to the liquid crystal order parameters.

By symmetry, the nematic order-parameter field, $N$, couples to the fermion density 
quadrupole \cite{Oganesyan2001}
\begin{align}
\mathbf{Q}(\vec{r},t)=\frac{1}{k_F^2}\psi^\dagger(\vec{r},t)
\left( 
\begin{array}{cc}
	\partial_x^2- \partial_y^2 & 2 \partial_x \partial_y
	\\
	2 \partial_x \partial_y & -\partial_x^2+\partial_y^2
\label{eq:density_quadrupole}
\end{array}
\right)
\psi(\vec{r},t).
\end{align}
In 2D, since the rotational group is $SO(2)$, the 
density quadrupole can be defined in terms of a two-component real director field ({\it i.e.\/} a headless vector) or, in terms of complex field
\begin{eqnarray}
Q(\vec{r},t)&=&Q_{11}(\vec{r},t)+i Q_{12}(\vec{r},t)\nonumber \\
&=&\psi^\dagger(\vec{r},t) \frac{(\partial_x+i \partial_y)^2}{k_F^2} 
\psi(\vec{r},t).
\label{eq:Q}
\end{eqnarray}
Same as the nematic order parameter, $Q$ is also invariant under rotations by $\pi$.

The coupling between $Q$ and $N$ is
\begin{align}
-g_N \int\mathrm{d}\vec{r}\mathrm{d}t (Q^\dagger N+ h.c.).
\label{eq:g_N}
\end{align}
Here $g_N$ is a coupling constant. Again, the chiral symmetry of the system requires that the effective action depends only on the real part of $Q^\dagger N$, and that there is no dependence on the imaginary part, since it is a pseudo scalar. The tensor form of this coupling is shown in
Appendix \ref{app:sec:tensor}.
In what follows we choose the sign of $g_N$ to be negative, so that a
positive expectation value of the nematic order parameter $\bar{N}$
means a FS stretched along the $x$ direction and compressed in the
$y$ direction, as shown in Figs. \ref{fig:FSCDW}(a) and (b). 

The sign of $g_N$ alone is not important. What matters is the relative 
sign between $g_N$ and the coupling constant $g$ defined in Eq. 
\eqref{eq:s_n_coupling_complex}. Under a redefinition of $N$ becoming 
$-N$, both $g_N$ and $g$ change sign. If $g \times g_N>0$, $\vec{Q}_S$ prefers
the direction in which the FS is stretched, but when $g \times g_N<0$, it 
prefers the direction where the FS is compressed. In general, 
the sign of $g \times g_N$ is determined by microscopic details of the system
to which this model may apply. 

If $Q_S=2k_F$, very close to a nesting condition the curvature of the FS 
controls the CDW instability as it controls how singular the charge susceptibility is near the nesting wave vector. In this case one finds  that it leads to the condition 
$g\times g_N<0$, when $Q_S$ connects two points on the Fermi 
surface where the curvature is smallest, as shown in Fig.\ref{fig:FSCDW}(b). 
In general, far from a nesting condition, the curvature of the FS alone is not the dominant 
factor, and the sign of 
$g\times g_N$ may be positive or negative, depending on the microscopic 
details.

The smectic order-parameter field should be coupled to the CDW of the 
fermions.  The CDW operator of the fermions, close to the ordering wave vector $Q_S$, is 
\begin{equation}
\displaystyle{n(\vec{q},\omega)= \int\frac{\mathrm{d}\vec{k}\mathrm{d}\Omega}{(2\pi)^3}\;
\psi^\dagger(\vec{k}+\vec{Q}_S+\vec{q},\Omega+\omega)  \;
\psi(\vec{k},\Omega)},
\label{eq:smectic-op-fermions}
\end{equation}
where $q\sim 0$. The smectic order-parameter field $\Phi$ couples to this 
fermion density wave 
$n$ as
\begin{align}
-g_S \int\mathrm{d}\vec{r}\mathrm{d}t (n^\dagger \Phi + h.c.).
\label{eq:g_S}
\end{align}
Integrating out the bosons, attractive four-fermion interactions are generated
of the form
\begin{align}
\frac{g_N^2}{\Delta_N} |Q|^2+\frac{g_S^2}{\Delta_S} |n|^2.
\end{align}
Hence, the order-parameter fields can be regarded as
Hubbard-Stratonovich fields used to decouple four-fermion interactions.
In this picture, the couplings between the order-parameter fields and fermions are 
measuring the strength of the attractive four-fermion term.

Gapless fermions will introduce nonanalytic terms to the low-energy effective theory of the nematics and 
smectics. For the case of the nematic order parameter, it was shown by Oganesyan and co-workers \cite{Oganesyan2001} that the 
fermions generate nonanalytic Landau damping terms\cite{Hertz1976,Millis1993},
so the theory of the isotropic-nematic metallic QCP becomes
\begin{align}
S_N=g_N^2 N(0)\int\frac{\mathrm{d}\vec{q}\mathrm{d}\omega}{(2\pi)^3}
\left(\frac{i |\omega|}{q}-\kappa_N q^2\right) N^{\dagger}(\vec{q},\omega)N(\vec{q},\omega),
\end{align}
The nematic susceptibility at this FL-nematic QCP is\cite{Oganesyan2001}
\begin{align}
\chi^N(\vec q,\omega)=&-i \avg{N^{\dagger}(\vec{q}, \omega) N(\vec{q},\omega)}_{\rm ret}
\nn\\
=&\frac{1}{\displaystyle{g_N^2 N(0)\left(\frac{i \omega}{q}-\kappa_N q^2\right)}}.
\end{align}

The phase mode of the nematic order-parameter field in the nematic phase, the nematic Goldstone mode, has an effective action of the form
\begin{eqnarray}
&&S_{\phi_N}=g_N^2 \bar{N}^2 N(0)\nonumber \\
&&\!\!\!\!\! \times \int \frac{\mathrm{d}\vec{q}\mathrm{d}\omega}{(2\pi)^3}
\;
\left(\frac{i |\omega|}{q} \sin^2 2\varphi_q -K_1 q_x^2-K_2 q_y^2 \right)\; |\phi_N(\vec{q},\omega)|^2
\nonumber \\
&&
\label{eq:quantum_nematic_goldstone}
\end{eqnarray}
where $\bar{N}$ is the expectation value of the nematic order parameter and $\varphi_q$ is the angle 
between $\vec{q}$ and the main axis direction of the nematic ordering. The stiffnesses $K_1$ and $K_2$ (the Frank constants) are given in Ref.\cite{Oganesyan2001}. 
With this action, it follows that the transverse nematic susceptibility in the electron nematic phase is\cite{Oganesyan2001}
\begin{align}
\chi^N_{\perp}(\vec q,\omega)=&-i\bar{N}^2 \avg{\phi_N (\vec{q}, \omega) \phi_N(-\vec{q},-\omega)}_{\rm ret}
\nn\\
=&\frac{1}{\displaystyle{g_N^2 N(0)\left(\frac{i \omega}{q} \sin^2 2\varphi_q -K_1 q_x^2-K_2 q_y^2\right)}}
\end{align}
For the case of a nematic order parameter aligned along the $x$-axis, the angular factor becomes $\sin^2 2\varphi_q=4 (q_x^2 q_y^2/q^4)$.

For the case of a charged smectic, a unidirectional CDW, a similar effect will be observed. Besides, if $Q_S$ connects to 
points on the FS which have just the opposite Fermi velocity 
as shown in Fig. \ref{fig:FSCDW}(b), the discontinuity leads
to another type of nonanalytic terms as will be shown in Sec.
\ref{sub:sec:Q_S_2kf}.

\section{the Nematic-Smectic Metallic Quantum Critical Point}
\label{sec:N_S_critical}

In this section, we study the metallic nematic-smectic QCP. The
two cases shown in Figs. \ref{fig:FSCDW}(a) $Q_S <2k_F$,  and (b) $Q_S=2k_F$, are studied separately.

Deep in the nematic phase, the amplitude fluctuations of the nematic order parameter are gapped, and the low-energy fluctuations are due to the 
nematic Goldstone mode, $\phi_N$, whose action is given in 
Eq. \eqref{eq:quantum_nematic_goldstone}. However, as the nematic-smectic QCP is approached (from the nematic side) the fluctuations of the smectic order parameter become progressively softer and, provided the quantum phase transition is continuous, become gapless at this QCP. In this scenario, the nematic phase looks like a ``fluctuating stripe'' phase qualitatively similar to the phenomenology of the cuprate superconductors, as discussed in Sec. \ref{sec:hts}.

The case $Q_S>2k_F$ will not be discussed here. The reason is that since now the CDW fluctuations with $Q_S>2k_F$ cannot decay into particle-hole pairs, in this case fermions only renormalize the coefficients of the smectic effective action, while the nematic fluctuations will still be Landau damped. For isotropic systems and  for $Q_S>2k_F$, the CDW (Lindhard) susceptibility $\chi(Q_S)$ in general decreases faster than linear as $\chi(2k_F)-A \; \sqrt{Q_S-2k_F}$, where $A$ is a constant. This implies that a CDW with $Q_S>2k_f$ is unlikely to be realized as it would require an anomalously attractive interaction at a large $Q_S$. However, for a lattice system the phase fluctuations of the nematic mode $\phi_N$  get gapped by lattice anisotropies and in this case the fermions only yield the trivial effect of renormalizing the coefficients of the effective action at the CDW transition. 

\subsection{$Q_S<2 k_F$}
\label{sub:sec:qS<2kF}

For $Q_S<2 k_F$, the leading contribution to the effective action of the order parameter field, resulting from integrating out the fermions, has  the form
\begin{align}
g_S^2\; \int \displaystyle{\frac{\mathrm{d}\vec{q}\mathrm{d}\omega}{(2\pi)^3}} \; 
 \Pi(\vec{Q}_S+\vec{q},\omega) \; |\Phi(\vec{q},\omega)|^2,
\label{eq:integrating_fermion}
\end{align}
Here $\Pi(\vec{Q}_S+\vec{q},\omega)$
is the CDW susceptibility of the fermions, given by the fermion loop integral (bubble)
\begin{align}
\Pi(\vec{k},\omega)\!
=-\!\!\int\!\! \displaystyle{\frac{\mathrm{d}\vec{p}}{(2\pi)^2}} \;
\displaystyle{\frac{n_F[\epsilon(\vec{p}+\vec{k})]-n_F[\epsilon(\vec{p})]}
{\omega - \epsilon\left(\vec{p}+\vec{k}\right)+\epsilon\left(\vec{p}\right)
+i0^+\; \textrm{sign}(\omega)}},
\end{align}
where $n_F(k)$ is the Fermi-Dirac distribution function.

The static part of the fermion CDW susceptibility depends on the details of the dispersion
relation from way above the FS to the bottom of the band. However, since
$\Pi(\vec{k},\omega=0)$ is analytic for $k<2 k_F$, the static part, $\Pi(\vec{Q}_S+\vec{q},\omega=0)$, 
will not change the analytic structure of Eq. \eqref{eq:ns_action}, but just renormalize the 
coefficients, in particular the critical value of the coupling constant. The important contribution comes from the dynamical part, 
$\Pi(\vec{Q}_S+\vec{q},\omega)-\Pi(\vec{Q}_S+\vec{q},0)$. The singular contributions to this integral are dominated by the behavior of the integrand around the four points on the FS, which are connected by the ordering wave vector
$\vec{Q}_S$, as marked with black dots on Fig. \ref{fig:FSCDW}(a). If we expand the dispersion
relation of the fermions around these four points,
$\epsilon(\vec{q})=\pm v_x q_x\pm v_y q_y$, to leading order
we get a Landau damping contribution
\begin{align}
\Pi(\vec{q}+\vec{Q}_S,\omega)-\Pi(\vec{q}+\vec{Q}_S,0)=\frac{i |\omega| }{2 \pi v_x v_y},
\label{eq:Pi-linear-omega}
\end{align}
which is linear in $|\omega|$.
The formula above can be checked by taking the limit of $Q_S\ll 2k_F$ or 
$Q_S\lesssim 2k_F$. In these two regimes, the fermion loop integral can be computed 
by RPA without expanding the dispersion relations around the four points. After setting
$v_F=1$, for 
$Q_S\ll 2k_F$, one finds $\Pi(\vec{q}+\vec{Q}_S,\omega)=i N(0)|\omega|/Q_S$ 
with $N(0)$ being the density of states
and for 
$Q_S\lesssim 2k_F$, $\Pi(\vec{q}+\vec{Q}_S,\omega)=i|\omega|\sqrt{k_F}/(2\pi \sqrt{2k_F-Q_S})$, 
which can be reached by expanding Eq. \eqref{eq:fermi_smectic_goldstone}.
Both of them agree with the general formula given above.

The term linear in $\omega$ in the effective action for the smectic field $\Phi$ of Eq. \eqref{eq:Pi-linear-omega}, which is due to the contributions of the fermions,
dominates over the ``naive'' dynamical term proportional to $\omega^2$ of the phenomenological 
theory. We can  thus write an effective action for the electron nematic-smectic quantum phase transition of the form

\begin{widetext}
\begin{eqnarray}
S&=&\int \frac{\mathrm{d}\vec{q}\mathrm{d}\omega}{(2\pi)^3}\;
C_0 i |\omega| \; |\Phi(\vec{q},\omega)|^2
-\int \mathrm{d}\vec{r}\mathrm{d}t \left(
C_y |\partial_y \Phi|^2
+C_x |(\partial_x -i \frac{Q_S}{2} \phi_N)\Phi|^2
+\Delta_S |\Phi|^2 
+u_{S} |\Phi|^4\right)
\nn\\
&&+\int \frac{\mathrm{d}\vec{q}\mathrm{d}\omega}{(2\pi)^3}
\;
\left(\widetilde{K}_0 \frac{i |\omega|}{q} \sin^2 2\varphi_q
-\widetilde{K}_1 q_x^2-\widetilde{K}_2 q_y^2 \right)
 |\phi_N(\vec{q},\omega)|^2,
 \nonumber \\
 &&
 \label{eq:quantum_mcmillan}
\end{eqnarray}
\end{widetext}
where $C_0=g_s^2/(2\pi v_x v_y)$, $\widetilde{K}_0=g_N^2 \bar{N}^2 N(0)$ and 
$\widetilde{K}_{1,2}=g_N^2 \bar{N}^2 N(0)K_{1,2}$.
The point $\Delta_S=0$ and 
$u_S>0$ is the nematic-smectic critical point.
With the nonanalytic dynamical term, the dynamic critical exponent of the field
$\Phi$ becomes $z=2$, instead of $z=1$ as it would generally be in the absence of fermions (or, if the fermions were gapped as in the case of an insulator).

 The nematic Goldstone mode $\phi_N$ has a dynamic critical exponent 
$z=3$ \cite{Oganesyan2001}, larger than the $z=2$ exponent for the smectic fluctuations. Thus, the Goldstone mode of the nematic order parameter $\phi_N$ and the smectic  $\Phi$ fluctuate on very different 
energy scales, with $\phi_N$ being the low-energy mode. If we only focus on the asymptotic low-energy 
theory, we should integrate out the high-energy mode $\Phi$. This process will lead 
to an effective theory of $\phi_N$. In turn, the low-energy mode $\phi_N$ will 
mediate interactions of the field $\Phi$.
However, we will show by a scaling argument that in the case of the quantum metallic system the coupling between the smectic field and the nematic Goldstone 
mode is irrelevant.

The action of Eq.\eqref{eq:quantum_mcmillan} is invariant under a rescaling parametrized by a factor $b$
\begin{flalign}
&t\rightarrow b^{-3} t,& &\vec{r}\rightarrow b^{-1} \vec{r},&
\nn \\
&\Phi(\vec{r},t)\rightarrow b^{3/2} \Phi(b^{-1} \vec{r},b^{-3}t),& &C_0\rightarrow b^{-1} C_0,&
\nn\\
&C_{x,y}\rightarrow C_{x,y},& &Q_S \rightarrow b Q_S,&
\nn\\
&\Delta_S \rightarrow b^2\Delta_S,& &u_S \rightarrow b^{-1} u_S,&
\nn\\
&\widetilde{K}_{0,1,2}\rightarrow b^{3} \widetilde{K}_{0,1,2}.& & &
\label{eq:scaling1}
\end{flalign}
where $C_x$, $C_y$, $K_0$, $K_1$, and $K_2$ are the stiffness in Eq. \eqref{eq:quantum_mcmillan}.

When $\Delta_S=0$, at the tree level and in the long-wavelength regime, 
both the gauge-like ``coupling constant'' $Q_S$ and $\widetilde{K}_{0,1,2}$
scale to infinite, but the ratio $Q_S^2/\widetilde{K}_{0,1,2}$ scales to $0$
as a function of $b^{-1}$. This implies that the gauge-like coupling is irrelevant.
Quantum fluctuations may change the tree-level scaling behavior as we include
loop corrections. However, for large enough $\widetilde{K}_{0,1,2}$ or small enough $Q_S$,
the irrelevancy of the gauge-like coupling will not be changed. As a byproduct,
we notice that $C_0$ and $u_S$ scale to zero in the long-wavelength regime, which
means that these two terms are irrelevant also. 
However, we should keep in mind that these operators are actually dangerous irrelevant,
in the sense that $C_0$ is necessary to find the proper equal-time correlation 
function for $\Phi$ and $u_S$ is necessary for stability in the ordered phase, and they are only irrelevant at this QCP.

Notice that at the QCP there are  two critical modes: the amplitude of the CDW order parameter, which has $z=2$, and the transverse (Goldstone) mode of the nematic phase, which has $z=3$ (and it is clearly dominant at low enough energies). Thus, we also
need to check the scaling behavior of $t\rightarrow b^{-2} t$ and
$\vec{r}\rightarrow b^{-1} t$ for the high-energy mode. Under this rescaling,
\begin{flalign}
&\Phi(\vec{r},t)\rightarrow b \Phi(b^{-1}\vec{r}, b^{-2}t),& &C_{0,x,y}\rightarrow C_{0,x,y},
\nn\\
&Q_S \rightarrow b Q_S,& &\Delta_S \rightarrow b^2\Delta_S,
\nn\\
&u_S \rightarrow u_S,& &\widetilde{K}_{0}\rightarrow b^{3} \widetilde{K}_{0},
\nn\\
&\widetilde{K}_{1,2}\rightarrow b^{2} \widetilde{K}_{1,2}. & & 
\label{eq:scaling2}
\end{flalign}
At the critical point where $\Delta_S=0$, it can be seen that $Q_S^2/\widetilde{K}_{0}$
also scales to $0$ as $b^{-1}$ in the long-wavelength limit, which means, for the $z=2$ mode, the
gauge-like coupling is still irrelevant. 

These conclusions are confirmed by one-loop perturbation theory calculations, presented  in Appendix 
\ref{app:sec:critical}, where we show that integrating out $\Phi$ (or $\phi_N$)
does not change the action of $\phi_N$ (or $\Phi$). This is one of our main results.

\begin{figure}[t]
\begin{center}
\includegraphics[width=0.45\textwidth]{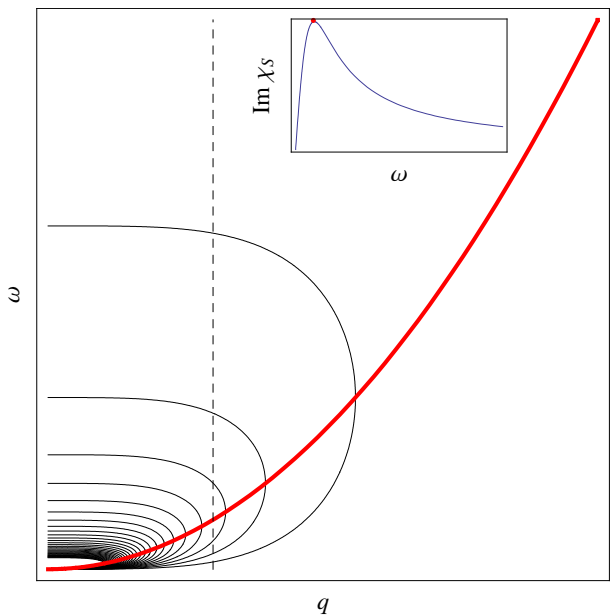}
\end{center}
\caption{(color online) The spectral density of the smectic susceptibility at the nematic-smectic QCP,
$Im \chi_S$, as a function of $q$ and $\omega$ for $Q_S<2k_F$. The spectral density is singular near the origin (lower right corner) and decays monotonically away from there. Here we show contour plots at constant spectral density with values from $0$ up to $2000$. The red line, $\omega=C_x q_x^2/C_0+C_y q_y^2/C_0$, marks the peak of the spectral density as a function of momentum $\vec q$ parallel to the nematic orientation. The inset is the energy dependence of $\textrm{Im} \chi_S$ at a fixed small momentum (along the dashed vertical line).
}\label{fig:critical_mode}
\end{figure} 

In conclusion, there are two essentially decoupled soft modes at the nematic-smectic
QCP. The nematic Goldstone mode is governed by the same action as in the nematic
phase, Eq. \eqref{eq:quantum_nematic_goldstone}. Since as the nematic-smectic QCP is approached from the nematic side, the nematic Goldstone mode and the smectic order parameters effectively decouple; the effective action for the smectic field $\Phi$ in this limit reduces to
\begin{align}
S_S=\int \frac{\mathrm{d}\vec{q}\mathrm{d}\omega}{(2\pi)^3}\;
 \left(i C_0 |\omega|-C_x q_x^2-C_y q_y^2-\Delta_S\right)\; |\Phi(\vec{q},\omega)|^2,
\end{align}
which implies that the dynamic smectic susceptibility is
\begin{align}
\chi^S=&-i \avg{\Phi^\dagger(\vec{q},\omega)\Phi(\vec{q},\omega)}_{\rm ret}
\nn\\
=&\frac{1}{i C_0 \; \omega-C_x q_x^2-C_y q_y^2-\Delta_S},
\label{eq:smectic-susceptibility}
\end{align}
where $\vec{q}$ is the momentum measured from the ordering wave vector $\vec{Q}_S$. Here $\Delta_S>0$ on the nematic side of this QCP. On the smectic (stripe) side of the quantum phase transition the $\Phi^4$ coupling, which is (dangerous) irrelevant at this QCP, cannot be ignored as it stabilizes the smectic ground state. The smectic susceptibility in the ordered smectic phase differs from that of Eq.\eqref{eq:smectic-susceptibility} in two standard ways: a) it acquires the usual delta function term peaked at the ordering wave vector, $\bar \Phi^2 \delta(\vec q)$, where $\bar \Phi$ is the expectation value of the smectic order parameter and $\vec q$ is measured from the ordering wave vector $\vec Q_S$, and b) the ``mass term''  ($\Delta_S$) in the denominator of the susceptibility becomes $2|\Delta_S|$. At the nematic-smectic QCP, $\Delta_S=0$, the smectic fluctuations are described by an overdamped critical mode with $z=2$. The spectral density of $\chi_S$ at the QCP is shown in Fig. \ref{fig:critical_mode}. It shows that most of the spectral weight is at small $\omega$ and $q$, which is typical for a critical mode, and the energy distribution curve at fixed momentum has a broad peak marked by the red line ($\omega=C_x q_x^2/C_0+C_y q_y^2/C_0$), indicating an overdamped critical mode with $z=2$.

The low-energy physics of the system will be dominated by $\phi_N$, the Goldstone mode of the nematic field,
whose behavior has  been studied extensively in Ref. \cite{Oganesyan2001,lawler-2006}. 
On the other hand, at higher energy scales
($\omega\sim q^2$), the effects of field $\Phi$, the amplitude mode of the smectic (or stripe) fluctuations, will become observable and, in this range, the system effectively has ``fluctuating stripes''. Notice that if the nematic Goldstone mode becomes gapped, say by the effects of the lattice, the smectic amplitude fluctuations become the only low-energy modes left. Finally, since the field $\Phi$ 
at the critical point has $z=2$, the effective total dimension of this theory is $4$. A standard Hertz-Millis type argument\cite{Hertz1976,Millis1993,Sachdev1999}  implies that in this case the $|\Phi|^4$ term is marginally irrelevant. Therefore, the 
Gaussian fixed point will have the correct scalings, up to logarithmic 
corrections. In particular, contrary to what happens in the classical case where this transition becomes weakly first order, at the nematic-smectic QCP the electron smectic susceptibility will acquire logarithmic corrections to scaling in the $ \omega \to 0$ and $\vec q \to 0$ limit, as can be deduced from standard arguments in classical and quantum critical phenomena\cite{amit-book,Sachdev1999}.

The gapless smectic amplitude mode $\Phi$, the {\em fluctuating stripe}, can be detected in inelastic light scattering experiments, much in the same way as in the case of conventional CDW materials. The existence of this mode also has observable effects on thermodynamic properties such as the 
low-temperature heat capacity. Since the dynamic exponent now is $z=2$, the fluctuation of the amplitude of the smectic mode leads to a $C\sim T$ contribution, which is subleading compared to the 
$C\sim T^{2/3}$ contribution of the nematic Goldstone mode $\phi_N$.

Besides the subleading contribution to the heat capacity, the critical fluctuations of the smectic 
order parameter also have an observable effect on the fermions as they profoundly change the character of these excitations. As pointed out 
by Oganesyan and coworkers \cite{Oganesyan2001}, the fermionic states in the 
nematic phase will become a non-FL due to the effects of the fluctuations of the nematic Goldstone
mode, an overdamped mode with dynamic critical exponent $z=3$. To leading order in perturbation theory, these authors found that for the most of the FS, the imaginary part of the fermion 
self-energy (the quasiparticle scattering rate) $\Sigma^{\prime\prime}(k_F,\omega)\sim|\omega|^{2/3}$,  which, as $\omega \to 0$ ({\it i.e.\/} as the FS is approached)
vanishes {\em slower} than $\omega$. Thus, in almost all of the FS the fermionic quasiparticle is no longer a well-defined state as the quasiparticle pole in the fermion Green function is lost.
This is the signature 
of a non-FL. However, for fermionic excitations propagating along the four main-axis directions of the nematic FS, the quasiparticle scattering rate now scales as  $\Sigma^{\prime\prime}(k_F,\omega)\sim |\omega|^{3/2}$. Although this is not the conventional $\omega^2$ behavior expected in an ordinary FL, nevertheless it is still consistent with the existence of a pole on the fermion spectral function, and a well-defined quasiparticle. Hence, in the nematic 
phase, except along these four special directions, the quasiparticles are not well defined. 

The $|\omega|^{2/3}$ behavior of the quasiparticle rate is in clear conflict with conventional FL behavior. It also implies that perturbation theory is breaking down in this system. Two approaches have been proposed to assess the non-perturbative behavior of the system. Using the non-perturbative approach of higher dimensional bosonization \cite{haldane94,Houghton93,Houghton00,castroNeto93,CastroNeto95}, Lawler and Fradkin showed that at the non-perturbative level the $|\omega|^{2/3}$ perturbative correction to the quasiparticle rate leads to a dramatic change in the behavior of the fermion propagator, which they found to have a vanishing quasiparticle residue and to exhibit a form of ``local quantum criticality" as it scales in frequency but not in momentum\cite{lawler-2006,lawler-2007}. On the other hand, Chubukov and Khveshchenko\cite{Chubukov06} used a resummed perturbation theory approach (on a similar problem) and argue that the $|\omega|^{2/3}$ behavior persists to all orders in perturbation theory. Although it is presently an open problem how to reconcile these two results, both analysis lead to the conclusion that the fermionic quasiparticles do not exist as well-defined excitations at the nematic-FL QCP and throughout the nematic phase (provided the nematic Goldstone modes remain gapless and overdamped).

At the nematic-smectic QCP, the fluctuations of the smectic field $\Phi$ ({\it i.e.\/} the ``fluctuating stripe'' mode) will also 
contribute to the fermion self-energy corrections 
(See Appendix \ref{app:sub:sec:n_s_critical_point} for details). For most points 
on the FS, the contribution to the quasiparticle rate $\Sigma^{\prime \prime}(k_F,\omega)$ of the fluctuations of the smectic field $\Phi$ 
is proportional to $\omega^2$, which is consistent with a conventional Landau behavior and a well-defined quasiparticle. However, at special points
on the FS satisfying $\epsilon(\vec k)=\epsilon(\vec k+\vec Q_S)$
(shown in Fig. \ref{fig:FSCDW}(a) as the four black dots), the contributions of the fluctuations of the smectic order-parameter field $\Phi$ to the quasiparticle rate scale as $\Sigma^{\prime \prime}(k_F,\omega) \sim |\omega|^{1/2}$. This $|\omega|^{1/2}$ behavior dominates even over the 
$|\omega|^{2/3}$ contribution of the nematic Goldstone mode. Hence, at the nematic-smectic  QCP the quasiparticle residue will vanish at these special points of the FS.
Note that these special points are precisely the positions on the FS where
FS reconstruction will take place due to the development of CDW order, as shown in Figs. \ref{fig:FSCDW} (a) and (c). Hence, it is not surprising 
to see that strong derivations from the Landau FL picture appear at these 
points at the QCP. This non-FL behavior is just the prelude of the  FS reconstruction in the ordered phase.

Finally, if the continuous rotational symmetry is broken explicitly by the anisotropic effects of the underlying  lattice (say through an anisotropic band structure) or by external
fields, at very low energies the nematic Goldstone mode of the nematic phase will have a finite  gap (generally $z=1$). If this gap is small enough, sufficiently close to the FS the fermion quasiparticle rate will show a crossover from the above mentioned $|\omega|^{2/3}$  above this gap to a conventional $\omega^2$ characteristic of a Landau FL. (Note that at the nematic-FL QCP the $|\omega|^{2/3}$ behavior is still obtained even for a lattice system\cite{dellanna-2006}.) Nevertheless, at 
the nematic-smectic QCP, the quantum critical fluctuations of the smectic mode still generate 
$|\omega|^{1/2}$ corrections to the quasiparticle rate of the fermions at the special points 
where the FS is going to be reconstructed.

\subsection{$Q_S=2k_F$}
\label{sub:sec:Q_S_2kf}

We now consider the special case of $Q_S=2k_F$. For the same reason as mentioned above, the coupling between the nematic 
Goldstone mode and the smectic field is irrelevant. Hence, at sufficiently low energies and close enough to the QCP, we can ignore their coupling and
consider the effective theory of the smectic order-parameter field alone. The effects of the irrelevant coupling to the nematic Goldstone mode can be put back in perturbatively {\em a posteriori}.

When $Q_S=2 k_F$, Eq. \eqref{eq:integrating_fermion} is still valid, although the structure of the fermion loop integral is very different. We can compute the effective theory of the bosonic modes by evaluating the fermion loop integral in 
the same way as we did in the previous section for the $Q_S<2k_F$ case. Again, we find that
$\Pi(\vec{Q}_S,0)$ depends on the microscopic details of the fermion dispersion relation, but $\Pi(\vec{Q}_S+\vec{q},\omega)-\Pi(\vec{Q}_S,\omega=0)$ are dominated by the behavior of the 
integrand around the two points on the FS connected by $Q_S$ as marked in 
Fig. \ref{fig:FSCDW}(b). We expand the dispersion relation around these two points as
$\epsilon=\mu \pm \delta k_y +\kappa \delta k_x^2/2$, where $\mu$ is the chemical potential,
$\kappa$ is the local curvature of the FS and $\vec{\delta k}$ is the momentum measured from
these two points. By evaluating the fermion loop integral using this dispersion relation, we 
can determine the low-energy Lagrangian density of the field $\Phi$ to quadratic order

\begin{widetext}
\begin{equation}
\displaystyle{\mathcal{L}_\Phi(\vec{q},\omega)=- \left( 
 \zeta \left\{\sqrt{q_y+\frac{\kappa q_x^2}{4}+\omega}\;
 +\sqrt{q_y+\frac{\kappa q_x^2}{4}-\omega}\; \right\}
 +\alpha_1 q_y+\alpha_2 q_x^2 \right) |\Phi(\vec q, \omega)|^2,}
 \label{eq:fermi_smectic_goldstone}
\end{equation}
which implies that the smectic susceptibility at the QCP of this case is
\begin{align}
\chi^S(\vec{q},\omega)=-i \avg{\Phi^\dagger(\vec{q},\omega)\Phi(\vec{q},\omega)}_{\rm ret}
=\displaystyle{
\frac{-1}
{\zeta \left\{\sqrt{q_y+\frac{\kappa q_x^2}{4}+\omega + i 0^+} +\sqrt{q_y+\frac{\kappa q_x^2}{4}-\omega - i 0^+}\; \right\}
 +\alpha_1 q_y+\alpha_2 q_x^2}}.
 \label{eq:susceptibilityPhi}
\end{align}
\end{widetext}

Here $\zeta=g_S^2/(2\pi\sqrt{\kappa})$ and $\alpha_{1,2}$ are determined by microscopic details.
Notice that $\zeta$ diverges as $\kappa$ vanishes, which
is related the fact that a flat FS has a logarithmically divergent CDW susceptibility.
In Sec. \ref{sec:inflection} we will show that at an inflection point of the FS, where $\kappa=0$, 
there is a stronger non-analytic behavior of the form $\sqrt[4]{q_y+\omega}$.

For this Lagrangian density, a naive dimension counting suggests the scaling dimensions $[q_x]=1$, $[q_y]=[\omega]=2$. Under this scaling, $\alpha_1$ and $\alpha_2$ are irrelevant. However, if we take a small $\omega$ expansion for $q_y+\kappa q_y^2/4<0$, the Lagrangian density becomes
\begin{align}
\mathcal{L}_\Phi=\left(\displaystyle{\frac{i \zeta |\omega|}{\displaystyle{\sqrt{|q_y+\frac{\kappa }{4} q_x^2}|}}}
-\alpha_1 q_y-\alpha_2 q_x^2 \right) |\Phi(\vec q, \omega)|^2,
\label{eq:2kfsmall}
\end{align}
since $\displaystyle{\sqrt{q_y+\kappa q_x^2/4+\omega}}$ and $\displaystyle{\sqrt{q_y+\kappa q_x^2/4-\omega}}$
cancel each other in the $\omega\rightarrow 0$ limit when $q_y+\kappa q_x^2/4<0$.
This Lagrangian density has scaling law $[q_x]=1$, $[q_y]=2$, and $[\omega]=3$. This scaling is only valid inside
the particle-hole continuum, while the naive scaling, $[q_x]=1$, $[q_y]=[\omega]=2$, is valid 
outside. This different behavior arises because $Q_S=2k_F$ is located at the edge
of the particle-hole continuum. Hence, the long wave length fluctuations may be inside or outside the 
particle-hole continuum, which leads to two possible different scaling behaviors.
Among these two different scaling behaviors, the $[q_x]=1$, $[q_y]=2$, and $[\omega]=3$ is the low 
energy mode in the long-wavelength limit. Hence, this mode dominates the low-energy physics and the scaling behavior.

The fact that $q_x$, $q_y$, and $\omega$ have different scaling dimensions 
is typical of anisotropic systems. For instance, in the smectic (stripe) phase of the quantum Hall 
state,  a similar scaling was found in Ref. \cite{barci-2002a}. 
(This scaling behavior was later on proved to be unstable \cite{Lawler2004}
due to the existence of an infinite set of marginal operators in that system.) Although the problem we are discussing here and the smectic quantum Hall state share the same scaling 
dimensions, they are actually quite different. In the case of the theory we are discussing in this section, it
is the theory of the QCP of the nematic-smectic transition, while in the quantum Hall 
case, the same scaling is found in the smectic phase. Second, the different scaling 
dimensions in the $x$ and $y$ directions in our case are due to the existence of the 
FS, which results in two different scaling dimensions depending on whether 
the momentum is perpendicular or parallel to the FS. In contrast, in the quantum Hall 
smectic phase, it is due to the residue symmetry of the broken rotational symmetry, which is 
the same as in the smectic phase of the classical liquid crystals. Third, the scaling dimension $3$ 
in the time direction is due to the nonanalytic dynamical term in the present problem
we are discussing, 
but in the quantum Hall smectic, it is due to the explicitly broken time-reversal symmetry. Most 
importantly, the quantum Hall smectic is essentially an insulator in the direction perpendicular 
to the stripes. In our case, however, the stripe has not yet formed at the QCP,
and the system is an anisotropic conductor in all directions. For our system, even inside the smectic 
phase, the FS is just partially gapped, so the conductivity in the direction 
perpendicular to the stripes is still non-zero. This difference is very important, since 
it means that in the quantum Hall smectic, the system is actually a $1D$ sliding Luttinger 
liquid, but in our case, the system is a full 2D structure. It is precisely the existence of the sliding
symmetry that makes the $[q_x]=1$, $[q_y]=2$, and $[\omega]=3$ scalings unstable for the 
quantum Hall smectic phase. For our system instead, due to the absence of the sliding symmetry, 
our $[q_x]=1$, $[q_y]=2$, and $[\omega]=3$ scalings will not experience the same 
instability as the quantum Hall smectic phase.

On the other had, non-Gaussian terms play an important role in our problem.
The leading non-quadratic term  is not the naive $|\Phi|^4$ term as in 
an ordinary $\Phi^4$ bosonic theory. Instead, the discontinuity at the FS 
introduces a non-analytic term $|\Phi|^{5/2}$, as shown in Appendix 
\ref{app:sec:phi_nonanalytic}. 
A similar non-analytical term has been found in the FL-ferromagnets transition by Maslov, Chubukov, and Saha \cite{Maslov2006}, where a non-analytic term $|m|^3$ in the thermodynamic potential, with $m$ being the ferromagnetic order parameter, results in non-analytic behaviors for the specific heat coefficient and the spin susceptibility. It is also reported in the same reference that this $|m|^3$ term will be replaced by
a term $\propto |m|^{7/2}$ at the QCP.
Our theory will have effective dimension $6$ at the Gaussian fixed point with scaling 
$[q_x]=1$, $[q_y]=2$, and $[\omega]=3$, where $|\Phi|^4$ is irrelevant ($[\Phi]=2$), but $|\Phi|^{5/2}$ is relevant. This relevant $|\Phi|^{5/2}$ term changes the scaling behavior of the critical theory away from the Gaussian theory. 

Hence, in contrast with the case 
$Q_S<2 k_F$, the quantum critical theory at $Q_S=2 k_F$ is not controlled
by the Gaussian fixed point.
This conclusion agrees with the results of the ``small momenta'' regimes 
($Q_S=2k_F$ incommensurate CDW critical point) of the theory of a CDW-FL QCP of Altshuler {\it et al} \cite{Altshuler1995},
who also found a relevant perturbation at the Gaussian fixed point and hence a runaway RG flow. 
Therefore, the Gaussian theory fails.
As a result, the irrelevant terms, which we
ignored in this study, will determine the fate of this 
transition. In general, there are two possible situations. If the runaway RG
flow has a (non-perturbative) discontinuity fixed point, the phase transition will become 
first order. This scenario is known as the ``fluctuation-driven first order transition''. 
However, there is also the possibility that the runaway RG flow has a non-trivial (and also non-perturbative)
quantum critical fixed point, in which case the transition is still second order but 
has a different scaling. It may even be possible to change from a first-order 
transition to a second one by tuning some control parameters and going through 
a quantum tricritical point. In any case, although it is generally assumed that a fluctuation-induced first-order transition normally results, actually it is not possible to determine which one of these two scenarios actually holds by the perturbative arguments we  are using here (and in Ref.\cite{Altshuler1995}).

\subsection{Commensurate CDW on a lattice}
\label{sec:commensurate-cdw}

Our analysis can be used for the case of a commensurate CDW QCP as well, 
{\it i.e.\/} the ``large momenta'' regime of Ref. 
\cite{Altshuler1995}. In this case, the Gaussian part of the effective
Lagrangian density will become
\begin{widetext}
\begin{align}
\mathcal{L}_\Phi&=&-\zeta \left(\displaystyle{
\sqrt{q_y+\frac{\kappa q_x^2}{4}+\omega}
+\sqrt{q_y+\frac{\kappa q_x^2}{4}-\omega}+\sqrt{-q_y+\frac{\kappa q_x^2}{4}+\omega}
+\sqrt{-q_y+\frac{\kappa q_x^2}{4}-\omega}}\right)
|\Phi(\vec q, \omega)|^2,
\end{align}
which leads to the smectic susceptibility
\begin{eqnarray}
\chi^S(\vec{q},\omega)&=&-i\avg{\Phi^\dagger(\vec{q},\omega)\Phi(\vec{q},\omega)}_{\rm ret}\nonumber \\
&=&\displaystyle{\frac{-1/\zeta}
{\sqrt{q_y+\frac{\kappa q_x^2}{4}+\omega +i 0^+}
+\sqrt{q_y+\frac{\kappa q_x^2}{4}-\omega -i 0^+}+\sqrt{-q_y+\frac{\kappa q_x^2}{4}+\omega + i0^+}
+\sqrt{-q_y+\frac{\kappa q_x^2}{4}-\omega -i0^+}}}.\nonumber\\
&&
\end{eqnarray}
\end{widetext}
The last two terms in the Lagrangian density appear, due to the mirror effect between momentum $\vec{k}$ and 
$\vec{G}-\vec{k}$ introduced by the band structure, where $\vec{G}=2\vec{Q}_S$ 
is a reciprocal lattice vector. 
The static part of the Lagrangian density does not have a full cancellation as the incommensurate case 
in Eq. \eqref{eq:2kfsmall} due to the fact that $\pm q_y+\kappa q_x^2/4$ can not be negative at the same time. Hence,
the naive scaling law, $[q_x]=1$, $[q_y]=2$, 
and $[\omega]=2$, is always valid,
which makes the effective dimension of this theory $5$. At the tree level, since $\Phi$ has dimension $2$,  $|\Phi|^{5/2}$ is marginal. In Ref. \cite{Altshuler1995}, it is shown that in this regime the 
coupling between the CDW order parameter and the fermions is marginal at
the tree level and, at one-loop level it becomes marginal irrelevant. Therefore, 
the $|\Phi|^{5/2}$ term should be marginal irrelevant here too. We conclude that the quantum critical theory of 
the commensurate CDW critical point is Gaussian, with logarithmic corrections to scaling.

In this case, the contribution of the quantum critical smectic field to the low-temperature specific heat
is proportional to $T^{3/2}$. The quasiparticle scattering rate $\Sigma^{\prime \prime}(k_F,\omega)$ acquires a contribution (due to the quantum smectic fluctuations) throughout most of the FS proportional to $\omega^{2}$ , which can be ignored in low-energies and it is consistent with a finite quasiparticle pole. However, for the fermionic excitations close to $\pm k_F \vec{e}_y$, $\Sigma^{\prime \prime}(k_F,\omega)  \sim |\omega|$, which suggests a marginal FL behavior. 
Details of this analysis can be found in Appendix \ref{app:sub:sec:n_s_critical_point}.
The result $\Sigma^{\prime \prime}(k_F,\omega)\sim |\omega|$ for $\vec{k}=\pm k_F \vec{e}_y$ agrees
with the findings of Ref. \cite{Altshuler1995}.

\subsection{The special case of a FS with inflection points}
\label{sec:inflection}

\begin{figure}[t]
\begin{center}
\subfigure[]{\includegraphics[width=0.2\textwidth]{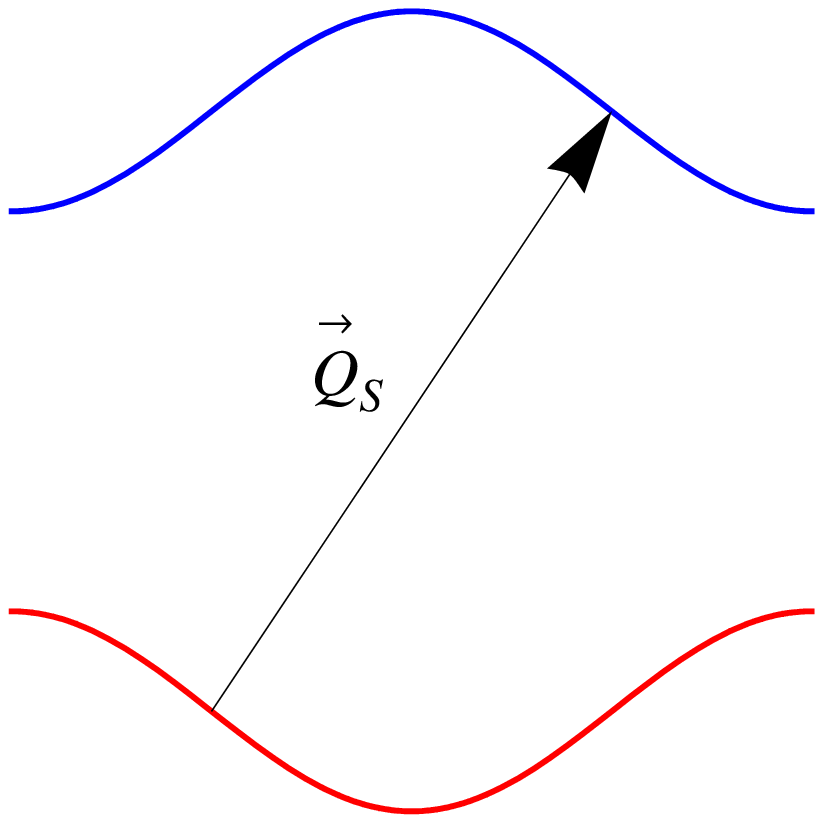}}
\subfigure[]{\includegraphics[width=0.2\textwidth]{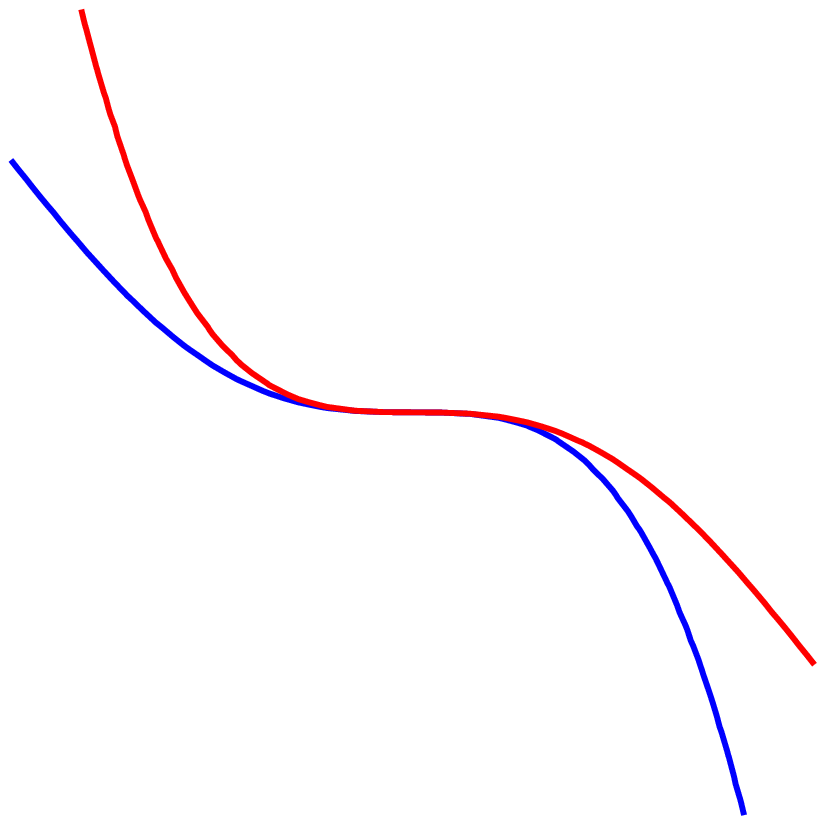}}
\end{center}
\caption{(online color) The FS near the inflection points.
(a) a FS with $4$ inflection points and
only the two connected by $\vec{Q}_S$ will be considered; (b)
the FS near the two inflection points by shifting
them together and rotated.
}\label{fig:inflection}
\end{figure} 

In the case of a two-dimensional system the FS is a curve. The point on the FS where the curvature vanishes is geometrically
an inflection point (Fig. \ref{fig:inflection}). In the $Q_S=2k_F$ case, if 
the two points connected
by $Q_S$ happen to be two inflection points, the scaling behavior will
become even more non-analytic than the cases we discussed above. Very close to the FS, we can assume that near the inflection 
points the dispersion relation of the fermions has an expansion of the form
\begin{align}
	\epsilon(\vec{q}+\vec{Q}_S/2)=\mu+q_y + a q_x^3+ b q_x^4+\ldots
	\label{eq:inf_disp}
\end{align}
Here $\mu$ is the chemical potential, the wave vector $\vec{q}$ has components
$q_x$ and $q_y$, which are the components perpendicular and tangent to the FS,  all
measured from the inflection point, and $a$ and $b$ are two constants. Assuming that the band structure has inversion symmetry, to obtain the dispersion relation around the other inflection point, 
we just need to reverse the vector $\vec{q}$.
The term quadratic in $q_x$ and $q_y$ is set to zero 
since the curvature of the FS vanishes at the inflection point. The quartic term of 
$q_x$ controls the FS nesting. Hence, if we decrease the coefficient
$b$ to zero, the CDW susceptibility will diverge, and a phase transition to a CDW state will be accessed
by tuning the parameter $b$. If we start from a nematic phase,
$b$ can be used to control the nematic-smectic transition.

The Gaussian terms in the effective Lagrangian density of the smectic order parameter field is
\begin{widetext}

\begin{equation}
\mathcal{L}_\Phi=-\gamma \left(\sqrt[4]{q_y+\frac{a q_x^3}{4}+\frac{b q_x^4}{8}+\omega}
+\sqrt[4]{q_y+\frac{a q_x^3}{4}+\frac{b q_x^4}{8}-\omega}\right)  |\Phi(\vec q, \omega)|^2,
\end{equation}
which implies the smectic susceptibility
\begin{align}
\chi^S(\vec{q},\omega)=-i\avg{\Phi^\dagger(\vec{q},\omega)\Phi(\vec{q},\omega)}_{\rm ret}
=\frac{-1/\gamma}
{\displaystyle{\sqrt[4]{q_y+\frac{a q_x^3}{4}+\frac{b q_x^4}{8}+\omega + i0^+}+\sqrt[4]{q_y+\frac{a q_x^3}{4}+\frac{b q_x^4}{8}-\omega -i 0^+}}}.
\end{align}

\end{widetext}
Here the constant $\gamma=(2/b)^{1/4}/(2\pi)$.
As we just mentioned, $b=0$ implies a nested FS and this is the reason that $\gamma$ diverges
as $b\rightarrow 0$.
In contrast to the ordinary $Q_S=2k_F$ incommensurate case studied before, there is no cancellation inside the particle-hole continuum for the inflection points, so that the naive scaling law, $[q_x]=1$,$[q_y]=3$, and $[\omega]=3$, is always valid.
The non-Gaussian terms now start at the order $|\Phi|^{9/4}$ as shown in Appendix
\ref{app:sec:phi_nonanalytic}. This term is irrelevant at tree level.

The low-energy heat capacity contributed by this mode has $C\sim T^{4/3}$.
The fermions near the inflection points have $\Sigma^{\prime \prime}\sim |\omega|^{13/12}$, which
means the fermionic quasiparticles near the inflection point will remain well defined,
even when the nematic-smectic QCP is reached.

\section{The Electronic Smectic Phase}
\label{sec:smectic_phase}

In the electronic smectic phase, {\it i.e.\/} in a conducting stripe phase, the electronic structure of the FL  quasiparticles is changed by the modulation imposed by the smectic order parameter, and an energy gap will develop at the locus of the former FS. As a result new electronic bands along the direction of the ordering wave vector will be formed, with a gap in the electronic energy spectrum $\Delta_S \propto g_S \Phi$.
As shown in Fig. \ref{fig:FSCDW}(c), if $2 k_F/Q_S$ is not close to an integer value, the highest band
will have a closed FS, {\it i.e.\/} an electron (or hole) pocket, while the lower band (or bands) has instead an open FS (an ``open orbit'').
For $2 k_F/Q_S$ close to an integer, as shown in Fig. \ref{fig:FSCDW}(d), there will only be an open FS. The reconstruction of the FS of the effective quasiparticles of a stripe phase in the cuprate {\hts} has been discussed recently\cite{millis-2007} as a possible explanation of the observation of Shubnikov-deHaas and deHaas-van Alphen oscillations  at relatively high magnetic fields\cite{doiron-leyraud-2007}.

We divide our discussion into two cases of interest: a) a smectic state in the continuum and b) a smectic state on a two-dimensional lattice.

\subsection{The electronic smectic phase  in a continuum system}
\label{sub:sec:con_smectic}

We consider first a system with continuous rotational and translational symmetries which are partially and spontaneously broken by the smectic order parameter.
We consider only the collective modes of the smectic order. 
The nematic collective modes will be added {\it a posteriori} using symmetry considerations. In this section we derive the effective order-parameter theory for the electronic smectic. The details of the calculation are presented  in Appendix 
\ref{app:sec:gs_mode_smectic}.

Let us consider a configuration of the complex smectic order-parameter field $\Phi$, which is a small deformation of the ground-state configuration with expectation value $\bar \Phi$,
\begin{align}
\Phi=\bar{\Phi}\left(1+\frac{\delta \Phi}{|\bar{\Phi}|}\right) e^{i \phi_\Phi}.
\label{eq:phi_fluctuation}
\end{align}
Here
$\delta \Phi$ measures the amplitude fluctuations of $\bar\Phi$ and 
$\phi_\Phi$ measures the phase fluctuations. 

At the mean-field level, {\it i.e.\/} ignoring fluctuations, the main effect of  the presence of a non-vanishing $\bar \Phi$ is that the original FS becomes folded along the direction determined by the ordering wave vector and a more complex electronic band structure results.
 The inter-band and the intra-band scatterings 
of the fermions will have different contributions to the effective low-energy theory.
For the smectic Goldstone mode, $\phi_\Phi$, in the static limit, the intra-band 
scatterings have no contributions but the inter-band scatterings will give a 
constant, which will cancel the constant coming from the terms 
$\Delta_S |\Phi|^2+ u_S |\Phi|^4$ for $Q_S<2k_F$, 
or $\Delta_S |\Phi|^2+ u_S |\Phi|^{5/2}$ for $Q_S=2k_F$,
provided the expectation value of $\bar \Phi$ satisfies the self-consistency equation of the 
mean-field theory. This exact cancellation of the constant term for the Goldstone 
mode is required by the Ward identity for translations along the ordering wave vector. The amplitude mode however will not have this 
cancellation, and a term proportional to $\delta \Phi^2$ will appear in its effective action.

For $\omega \neq 0$ and $\vec q \neq \vec 0$, the intra-band and inter-band 
scatterings also yield different contributions. The effective low-energy Lagrangian 
density of the smectic Goldstone mode $\phi_{\Phi}$ (a real field in position space and time) takes the form (see Appendix \ref{app:sec:gs_mode_smectic})
\begin{widetext}
\begin{equation}
\mathcal{L}_{\phi_\Phi}=g_S^2|\bar{\Phi}|^2 N_S(0)
\left(B(\varphi_q)\frac{\omega^2}{k_F g_S |\bar \Phi|}+i A(\varphi_q) 
\frac{|\omega| q}{k_F g_S |\bar \Phi|}-\kappa_S(\varphi_q) q^2 \right) |\phi_\Phi(\vec q, \omega)|^2,
\label{eq:CDW_Goldstone}
\end{equation}
\end{widetext}
where $\varphi_q$ is the angle between $\vec{q}$ and the stripe direction and $N_S(0)$
is the density of states in the smectic phase.
$A(\varphi_q)$, $B(\varphi_q)$, and $\kappa_S(\varphi_q)$ are coupling constants that depend on 
microscopic details and the direction of $\vec{q}$, which reflects the anisotropic 
nature of the smectic phase. We will neglect the
direction dependence of these three coefficients since
they result in irrelevant contributions at low energies and long distances. The first term $\propto\omega^2$ in Eq. \eqref{eq:CDW_Goldstone} is due to inter-band scattering, while
the second, $\propto i |\omega|q $, is due to intra-band scattering. The intra-band
contribution does not have the typical $i|\omega|/q$ form, because the contributions from the 
intra-band scattering vanish in the $q=0$ limit (see Appendix 
\ref{app:sec:gs_mode_smectic}.) The dynamic critical exponent is $z=2$ at the QCP but $z=1$ in the smectic phase. This discontinuity is reflected in the singularity of Eq. \eqref{eq:CDW_Goldstone} at $\bar{\Phi}=0$.

This behavior of the smectic Goldstone mode is very similar to that of the spin Goldstone mode in the commensurate antiferromagnetic 
phase (commensurate SDW) of a 2D FL studied by Sachdev, Chubukov, and Sokol \cite{Sachdev1995}. In fact, although they are studying the commensurate SDW, while we are studying incommensurate CDW,
if the nematic fluctuations in our system are gapped by a lattice background or external fields which breaks the continuous rotational symmetry, the nematic-smectic transition we studied here shares many common features with their work (especially ``type B'' in Ref. \cite{Sachdev1995}). Both these two transitions are driven by FL instabilities with a finite ordering wave vector and fermions provide similar non-analytic damping terms for low-energy bosonic excitations in both systems. For both cases, the QCP has $z=2$ and the ordered phase has a Goldstone mode with $z=1$.

An RPA calculation (see also Appendix \ref{app:sec:gs_mode_smectic}) shows that the amplitude mode fluctuations $\delta \Phi$ of the electronic smectic order parameter has an effective low-energy Lagrangian density of the form
\begin{align}
\mathcal{L}_{\delta \Phi}=g_S^2 N_S(0) \left(i\frac{|\omega|}{q}- |\Delta_S| \right) |\delta \Phi(\vec q, \omega)|^2.
\end{align}
The dynamical term $i|\omega|/q$ comes from the intra-band scattering. In contrast with the case of the Goldstone mode $\phi_\Phi$, there is no cancellation of the intra-band contributions
at $q=0$ for the amplitude mode $\delta \Phi$. Inter-band scattering contributes an irrelevant 
$\sim \omega^2$ dynamical term which is ignored. 
The longitudinal CDW susceptibility in the smectic ordered phase is
\begin{align}
\chi^S_{\parallel}(\vec{q},\omega)=&-i\avg{\delta \Phi(\vec{q},\omega)\delta \Phi(-\vec{q},-\omega)}_{\rm ret}
\nn\\
=&\displaystyle{\frac{1}{g_S^2 N_S(0) \left(\displaystyle{\frac{i \omega}{q}}- |\Delta_S|\right)}},
\end{align}
where $\vec{q}$ is measured from the ordering wave vector $\vec{Q}_S$

Hence, in the electron smectic phase both the phase and amplitude modes of the smectic order parameter scale with a dynamical exponent $z=1$. This result is interesting, because in most other examples of continuous symmetry breaking, the amplitude mode is either 
gapped or has a smaller dynamic critical 
exponent $z$ than the Goldstone mode. As a result, at lower 
energies, the amplitude mode will become weaker, and can be ignored from the 
effective low-energy theory. However, in this case, since both the amplitude 
mode and the Goldstone mode have the same dynamic critical exponent, no matter how 
low the energy scale is these two modes can no longer be separated.

Nevertheless, in our problem, the amplitude mode is irrelevant for the following reasons. 
Using Eqs. \eqref{eq:g_S} and \eqref{eq:phi_fluctuation}, we see that
both the amplitude CDW fluctuations $\delta \Phi$ and the CDW Goldstone modes $\phi_{\Phi}$ couple to fermionic CDW operator $n(\vec q,\omega)$ in the 
form: 
\begin{eqnarray}
&&g_S \; n(\vec q,\omega)\; \delta \Phi (-\vec q,-\omega)+h.c., \nonumber \\
&&i g_S\;  \bar\Phi \; n(\vec q,\omega)\;  \phi_\Phi(-\vec q,-\omega)+h.c.,
\end{eqnarray}
but the coefficients in the action of $\delta \Phi$ ($g_S^2N_S(0)$) and $\phi_\Phi$ 
($g_S|\bar{\Phi}|N_S(0)A/k_F$, $g_S|\bar{\Phi}|N_S(0)B/k_F$ and $g_S^2|\bar{\Phi}|^2 N_S(0)\kappa_S$)
have different scaling behaviors. With $z=1$, the coefficients for $\delta \Phi$ have dimension
$3$, but the coefficients $\phi_\Phi$'s coefficients have dimension $1$. As a result, the coupling between the smectic Goldstone mode 
$\phi_\Phi$ and the fermions is actually more relevant than the coupling to the amplitude mode $\delta \Phi$. Hence, we will not consider 
$\delta \Phi$ in what follows.

We finally  need to couple the smectic Goldstone mode, $\phi_\Phi$, with the nematic 
Goldstone mode $\phi_N$. Similar to the classical case, $\phi_N$ couples to the electronic smectic field $\Phi$ as a gauge field (Eq. \eqref{eq:ns_action}). 
Hence, in the electron smectic phase, upon substituting Eq. \eqref{eq:phi_fluctuation} into 
Eq. \eqref{eq:ns_action}, the nematic Goldstone mode $\phi_N$ develops a mass term $\sim|\bar \Phi|^2 \phi_N^2$, {\it i.e.\/} it acquires a gap.
As in the  classical case, the fluctuations of the nematic order produce bending of the smectic order, 
which  costs energy. Thus, here too, this effect leads to the expulsion of the nematic Goldstone mode.

Since  the dynamics of the nematic Goldstone mode $\phi_N$ has the structure $i|\omega|/q$, the mass term $|\bar \Phi|^2 \phi_N^2$ of the nematic Goldstone mode in the smectic phase
makes this mode to have $z=1$ dynamics, just as the smectic Goldstone mode $\phi_\Phi$ has. However, if we are 
close to the QCP, where the smectic amplitude fluctuation $\bar \Phi$ is small, the energy scale of the smectic Goldstone mode 
$\phi_\Phi$ will be lower than that of $\phi_N$. Hence, upon integrating out the nematic Goldstone mode $\phi_N$, we obtain an effective low-energy action for $\phi_\Phi$ of the form
\begin{widetext}
\begin{equation}
\mathcal{L}_{\phi_\Phi}=g_S^2|\bar{\Phi}|^2 N_S(0) 
\left(B\frac{\omega^2}{k_F g_S |\bar \Phi|}
+i A\frac{|\omega| \sqrt{\kappa_1  \kappa_2^{-1} q_x^4+ q_y^2}}{k_F g_S|\bar \Phi|}
-\kappa_1 q_x^4-\kappa_2 q_y^2 \right) \; |\phi_\Phi(\vec q, \omega)|^2.
\label{eq:smectic_CDW_no_lattice}
\end{equation}
\end{widetext}
where $A$, $B$ are two coefficients, and $\kappa_1$ and $\kappa_2$ are the  defined in Eqs. \eqref{eq:class_smectic_goldstone} and \eqref{eq:kappas}.
Here, as in the classical case, the coefficient of the $q_x^2$ term vanishes by rotational invariance.

By inspection of Eq. \eqref{eq:smectic_CDW_no_lattice} we see that the scaling dimensions now are $[q_x]=1$ and $[q_y]=[\omega]=2$.  
Hence, we can conclude that the fluctuations of $\phi_\Phi$ contributes to the low-temperature specific heat which scales with temperature as $C\sim T^{3/2}$.
It follows from this scaling analysis that in this case the  system is
above its upper critical dimension. Hence, true long-range order 
exists.  The logarithmic corrections generated by the higher-order terms in classical smectics
\cite{Grinstein1981} will not be present.
Notice that these scaling dimensions are the same as what we got from the 
phenomenological theory in Sec. \ref{sec:classical}, before it was coupled  
to the fermions. However, the physics is now very different. Most important of all, in the presence of 
fermions, the smectic Goldstone mode is damped, which is not the case in the 
phenomenological theory.

A  one-loop calculation of the fermion self-energy shows that the fermionic quasiparticles in the electronic smectic phase is a FL (Appendix \ref{app:sub:sec:smectic_phase}). Indeed, for most of the FS, the quasiparticle scattering rate at low frequency is $\Sigma^{\prime \prime}(k_F,\omega) \sim \omega^2\log |\omega|$. At the 
special points on the FS where the Fermi velocity is parallel to the ordering wave vector $\vec{Q}_S$ ({\it i.e.\/} perpendicular to the stripes) ( marked in Fig. \ref{fig:FSCDW}(c)), $\Sigma^{\prime \prime}(k_F,\omega)\sim |\omega|^{3/2}$ for small $\omega$. However, 
this $|\omega|^{3/2}$ scaling will not be observed for $2k_F/Q_S$ close to an integer, since no point on the FS will have $\vec{v}_F \parallel \vec{Q}_S$ as shown in Fig.  \ref{fig:FSCDW}(d).

\subsection{The electronic smectic phase on a lattice}
\label{sub:sec:lattice_smectic}

For most physical electronic systems the continuous rotational symmetry is reduced to a discrete point group symmetry of the underlying lattice.  As it is well known\cite{Kivelson1998,Oganesyan2001} the explicit breaking of rotational invariance by the lattice has important consequences for an electron nematic state. For the simple square lattice, the $O(2)/\mathbb{Z}_2$ symmetry of the continuum nematic state reduces to an Ising-like  $\mathbb{Z}_2$ symmetry, an ``Ising nematic.'' On the other hand, at the nematic-FL QCP symmetry breaking by the underlying lattice are irrelevant\cite{dellanna-2006}, provided the quantum phase transition remains continuous which, in many instances does not appear to be the case\cite{kee-2003,Khavkine-2004}. Lattice effects manifest in the electronic structure and hence on the allowed shape of the FS of the FL state. Lattice anisotropies will then  break the continuous rotational symmetry and act as explicit symmetry breaking fields. In this case, the nematic phase does not have a true Goldstone boson which now becomes gapped. At the level of the effective theory this effect shows up by the presence of a term proportional to $\phi_N^2$ in the effective action of the nematic Goldstone mode, which is just an allowed mass term since there is no Ward identity to prevent it. This term makes $\phi_N$ a $z=1$
mode, in contrary to $z=3$ without lattice. Hence, the ``pseudo-Goldstone'' mode $\phi_N$ has now become irrelevant at low energies. This changes a number of things. For one, the non-FL behavior of the nematic phase in the continuum is replaced by an anisotropic FL state with well-defined quasiparticles. Naturally if the effects of the lattice are weak enough, at sufficiently high energies (or temperatures) they can be neglected and the non-FL effects become detectable above this crossover. On the other hand, as the quantum phase transition to the electronic smectic phase is approached the gapped nematic Goldstone modes become irrelevant and the low-energy physics is dominated instead by the fluctuations of the smectic mode, which in the continuum case is a higher-energy excitation. In other terms, one now obtains a ``fluctuating stripe'' regime close enough to the nematic-smectic QCP.

In the electronic smectic phase, the lattice anisotropy makes the nematic Goldstone mode irrelevant, which can now be neglected in the effective low-energy theory.
As a result, we only need to consider the smectic Goldstone mode to understand the 
low-energy physics of the smectic phase.
Depending on whether the smectic (CDW) is pinned down by the lattice or not, the physics of the smectic Goldstone mode
behaves very differently. In the absence of lattice pinning, the Lagrangian density shown in 
Eq. \eqref{eq:CDW_Goldstone} 
will be the proper low-energy theory of the smectic Goldstone fluctuations, so 
the scaling behavior is $[q_x]=[q_y]=[\omega]=1$ and the transverse smectic susceptibility is
\begin{widetext}
\begin{align}
\chi^S_{\perp}(\vec{q},\omega)=&-i \bar{\Phi}^2\avg{\phi_\Phi(\vec{q},\omega)\phi_\Phi(-\vec{q},-\omega)}_{\rm ret}
=\frac{1}{g_S^2 N_S(0)}
\displaystyle{\frac{1}{B(\varphi_q)\frac{\omega^2}{k_F g_S |\bar \Phi|}+i A(\varphi_q) 
\frac{\omega  q}{k_F g_S |\bar \Phi|}-\kappa_S(\varphi_q) q^2}},
\end{align}
\end{widetext}
where $\vec{q}$ is measured from the ordering wave vector $\vec{Q}_S$.
This low-energy behavior is different from the electronic smectic  in the continuum. If the 
lattice effects are weak enough, we expect a crossover with increasing energy scales  from the behavior of a
smectic coupled to the lattice to one without this coupling.

By following the same type of analysis used in the previous subsections we can conclude that 
the heat capacity has a contribution due to the fluctuations $\phi_\Phi$ with a  $T^2$ temperature dependence. To the one-loop level, 
the fermionic quasiparticles at the FS will have a scattering rate $\Sigma^{\prime \prime}(k_F,\omega) \sim \omega^2\log|\omega|$ for $\omega \to 0$ at most part of the FS, while  $\Sigma^{\prime \prime}(k_F,\omega) \sim \omega^2$
for some special points on the FS discussed in Appendix \ref{app:sub:sec:smectic_phase}.
This observation is consistent with a FL behavior (Appendix \ref{app:sub:sec:smectic_phase}).

For a pinned CDW, the cancellation of the intra-band scattering for $\phi_\Phi$ no longer holds. 
Therefore, the dynamical term is $\sim i|\omega|/q$ with $z=1$. The coupling between $\phi_\Phi$ and the fermions is irrelevant in the pinned smectic phase, and the FL picture will hold.

In general, there are two sources for pinning the CDW: lattice and impurities.
We will consider the lattice pinning first. Lattice pinning is relevant for commensurate 
CDWs. For incommensurate CDWs, however, some lattice distortions will be required to 
pin the CDWs. A distorted lattice can pin down an incommensurate CDW only if the CDW order parameter
is large enough. This is because the energy gain by pinning vanishes as 
$|\bar \Phi|$ goes to $0$. On the other hand, the energy needed to distort 
the lattice will not go to zero with decreasing $|\bar \Phi|$. 
For most conventional CDW materials, the CDW ordering is very strong at low $T$ so that
incommensurate CDWs will always be pinned down by distorted lattice
\cite{Frohlich1954,Fogle1972,Lee1974,Gruner1988} and unpinned CDW only
appears at finite $T$ above a phase transition \cite{McMillan1976}. 

However, for our systems, there is a second-order phase transition from nematic
to smectic. When we are close to the nematic-smectic QCP, the smectic order parameter
will be small, so that an unpinned incommensurate CDW phase is stable against a lattice 
background at least when it is close to this QCP. When we are far 
from the critical point, the smectic order parameter becomes large, so the energy gain 
by pinning may exceed the energy cost of distortion. Hence, a pinned CDW phase may form.

Throughout this paper we ignore the effects of  quenched disorder. Impurities, and more generally disorder, affect strongly all electronic liquid crystal phases as they couple linearly to the their order parameters, leading to the destruction of these ordered phases and to  glassy-type states\cite{Kivelson1998}. For the case of the CDW phases this problem was studied extensively in the 1980s, for which pinning is relevant for $d<4$ for systems with short-range interactions 
\cite{sham-1976,imry-1975} and $d<3$ for long-range Coulomb interactions \cite{bergman-1977}. In the case of these quantum phase transitions the effects of quenched disorder, even in the clean limit, are only poorly understood and we will not explore these problems in this work.

\section{Finite-Temperature Crossovers and Thermal Phase Transitions}
\label{sec:finiteT}

\begin{figure}[h!]
\begin{center}
\subfigure[~isotropic system]{\includegraphics[width=0.33\textwidth]{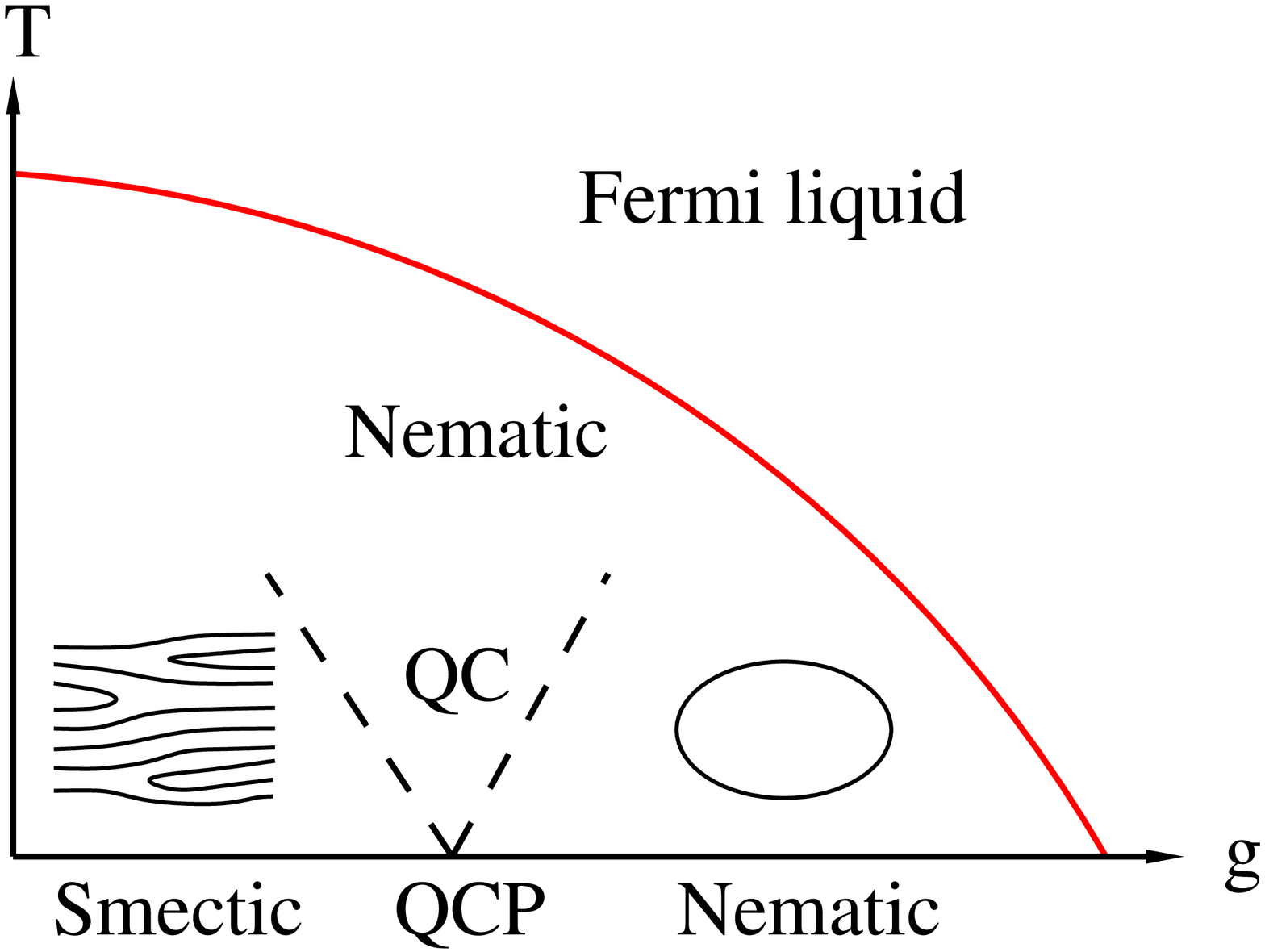}}
\subfigure[~with a lattice background]{\includegraphics[width=0.33\textwidth]{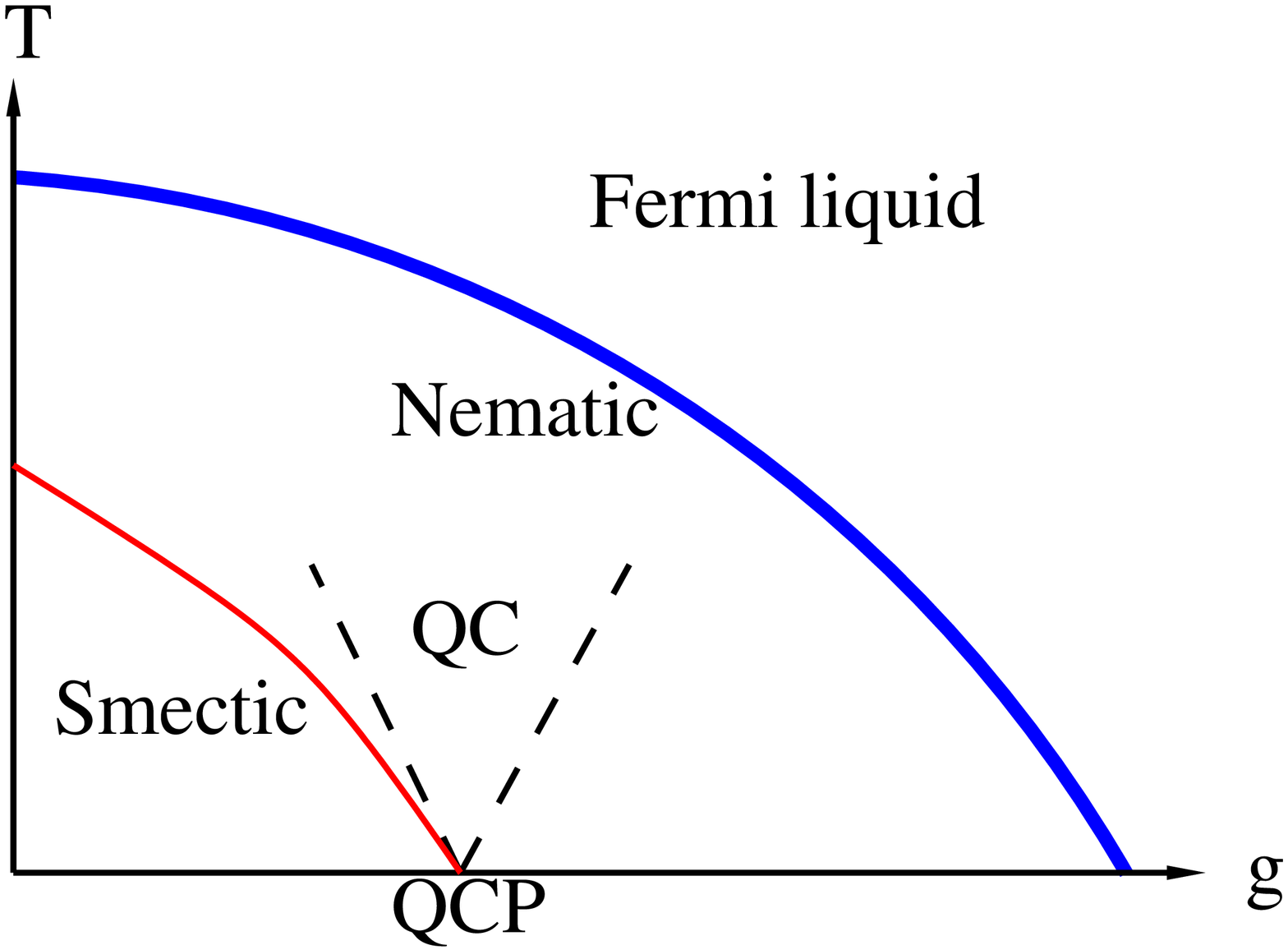}}
\subfigure[~pinned CDW]{\includegraphics[width=0.33\textwidth]{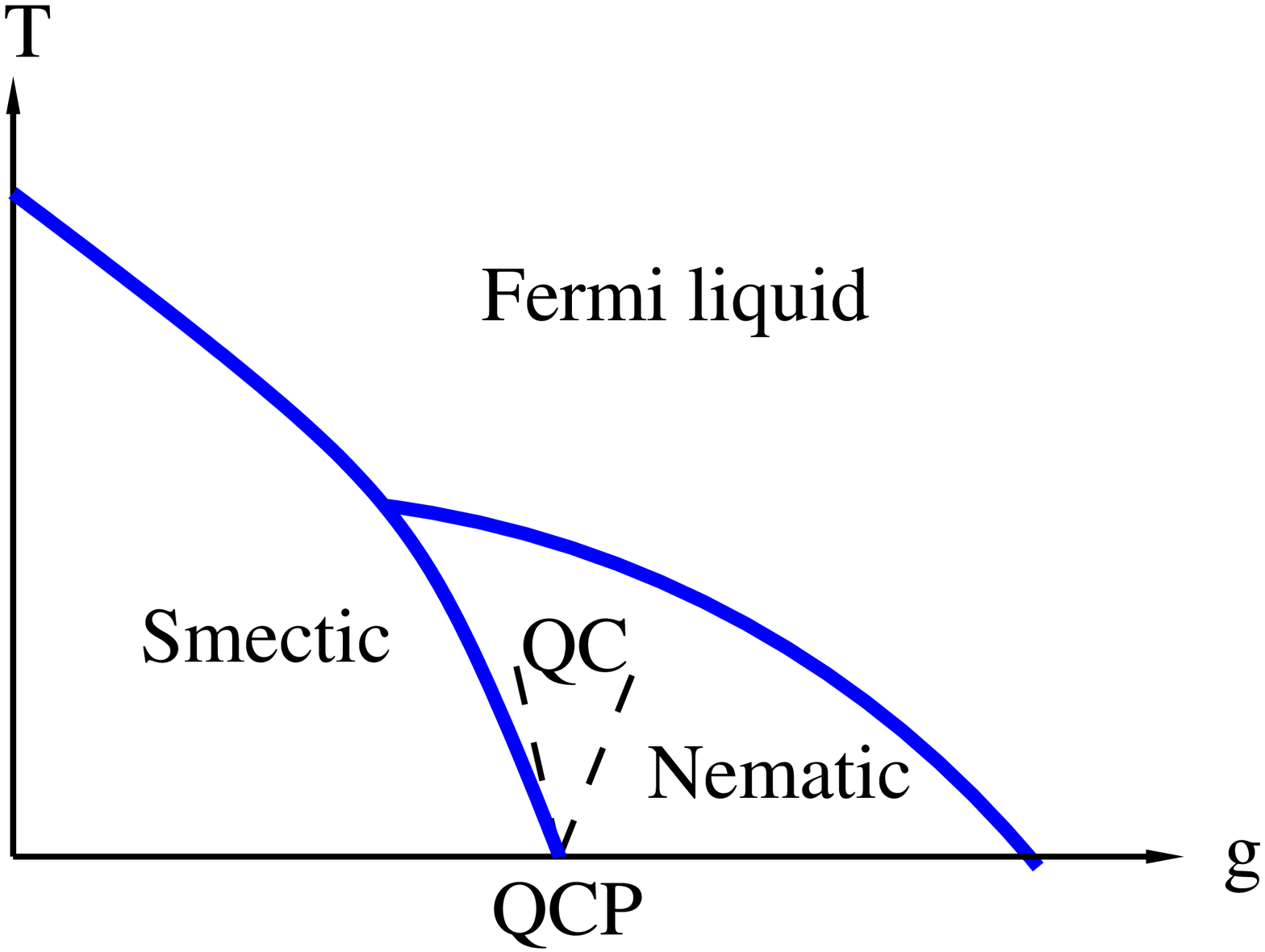}}
\end{center}
\caption{(online color) The schematic phase diagrams at finite $T$ for (a) an isotropic system, (b)  a system with
a lattice background, and (c) systems with a pinned CDW in the smectic phase. 
The horizontal axis is a control parameter that drives the quantum phase transitions: Fermi liquid $\rightarrow$ nematic $\rightarrow$ smectic at $T=0$. The vertical axis is the temperature $T$.
The thick (blue) lines are phase transitions belong to the Ising (or $q$-state Potts model) universality class while thin (red) lines are $KT$ transition phase boundaries. The dashed lines mark the crossover to the quantum critical regime (QC). In panel
(a) we show that the nematic phase (in the absence of lattice symmetry breaking) only has long range order at $T=0$, and it is a critical KT phase at all temperatures below the KT transition. In panel b) the lattice reduces the symmetry to $\mathbb{Z}_2$ and there is long-range Ising nematic order at finite temperatures. In this case  there is also a finite-temperature smectic phase which is critical (KT) if the smectic (CDW) order is unpinned and has long-range order in the latter case (shown in panel c).}
\label{fig:finiteT}
\end{figure}

We now discuss the effects of a finite-temperature on the electronic nematic and smectic phases and on their quantum phase transition. In Fig. \ref{fig:finiteT}, we present schematic phase diagrams for the three cases of interest: a) the isotropic case (no lattice), b) the system with a lattice background (a square lattice in this case) and an unpinned electron smectic (CDW) and in c) the case of a lattice with pinned CDW.

In the nematic phase and at the nematic-smectic QCP, the nematic Goldstone
modes have a finite temperature equal-imaginary time correlation function, {\it i.e.\/} the transverse nematic susceptibility,
\begin{align}
\chi^N_\perp(\vec{q})=&\bar{N}^2 \avg{\phi_N(\vec{q},\tau) \phi_N(-\vec{q},\tau)}_{\rm ret}
\nn\\
=&\frac{1}{\beta g_N^2 N(0)}\sum_{\omega_n} \frac{1}{\frac{|\omega_n|}{q}\sin^2 2
\varphi_q +K_1 q_x^2 +K_2 q_y^2}
\end{align}
where we have set $k_B$  to $1$ so $\beta=1/T$ and the sum runs over all bosonic
Matsubara frequencies, $\omega_n=2\pi n T$, where $n \in \mathbb{Z}$. The susceptibility
$\chi^N_\perp(\vec{q})$ is singular as $\vec q \to 0$ (due to the $\omega_n=0$ mode, where it takes the asymptotic form $\sim
T/ q^2$. This behavior suggests that the nematic order parameter field has power-law correlations at
finite $T$.
This conclusion agrees with the classical theory of a 2D nematic
phase, which belongs to
the Kosterlitz-Thouless (KT) universality class\cite{toner-1980,Chaikin98}.

By comparing the $T/q^2$ term with the equal-time correlation function
at $T=0$, which
is $\sim q \log(E_F/q)$ where $E_F$ is the Fermi energy, which enters
the calculation as a
high energy cutoff, we see that  $T=0$ behavior becomes dominant
when $q\gg T^{1/3}$.

From the theory of classical liquid crystals\cite{DeGennes1993,Chaikin98}, we know that in a fully translationally invariant system, {\it e.g.\/} in the absence of a lattice, there is no stable
finite-temperature smectic
phase in two space dimensions\cite{toner-1980}, which implies that the equal-time correlation function of the
smectic order parameter decays exponentially fast as a function of distance at any finite $T$. Due to the effects of the proliferating dislocations of the smectic the actual finite-temperature phase is a nematic.

In the nematic phase and at the nematic-smectic QCP, the finite $T$ equal-time
correlation function of the smectic fields, {\it i.e.\/} the smectic susceptibility, is
\begin{align}
\chi^S(\vec{q})=&\avg{\Phi(\vec{q},\tau) \Phi(-\vec{q},\tau)}_{\rm ret}
\nn\\
=&\frac{1}{\beta}\sum_{\omega_n} \frac{1}{C_0 \; |\omega_n|+C_x q_y^2 +C_y
q_y^2+\Delta}
\label{eq:finite_T_smectic}
\end{align}
Here, $\vec{q}$ is measured from the ordering wave vector $\vec{Q}_S$ and $\Delta=\Delta_S + f(T)$, where $f(T)$ is a function of temperature
which
vanishes at $T=0$. The leading term in $f(T)$ comes from the quartic
term of the
smectic field $\sim u_S T$ and the gauge-like couplings between the
nematic Goldstone
mode and the smectic fields $\sim Q_S^2 T$, where the ordering wave vector $Q_S$ represents the strength of
the gauge-like coupling. The absence of a finite-$T$ smectic-ordered phase implies that $\Delta \neq 0$ for $T>0$.
Hence, on the nematic
side, at low enough temperatures $\Delta>0$ and it is essentially equal to $\Delta_S>0$. However, in the quantum critical
region (denoted by QC in Fig. \ref{fig:finiteT}), 
the singular behavior of $\chi^S(\vec{x})$
is regulated purely by finite temperature. Notice that
$f(T)$ depends on irrelevant operators of the QCP; we conclude that the
infrared divergence is regulated by irrelevant operators, including the gauge-like coupling between smectic and the nematic Goldstone mode, which implies that this gauge-like coupling is a dangerous irrelevant term near the QCP and become relevant at finite $T$. This conclusion agrees with the theory of the classical nematic-smectic transition at finite $T$, where the gauge coupling is known to be relevant.
The boundary between the nematic and the quantum critical regime is determined by the condition
$f(T)\sim \Delta_S$, which is $T\sim \Delta_S$. 

On the smectic side, the equal-time correlation function of the smectic Goldstone
mode fluctuations, {\it i.e.\/} the transverse smectic susceptibility $\chi^S_\perp(\vec q)$ (where $\vec q$ is the momentum measured from the ordering wave vector $\vec Q_S$ of the smectic phase) is
\begin{widetext}
\begin{equation}
\chi^S_\perp(\vec q)\equiv |\bar{\Phi}|^2\avg{\phi_S(\vec{q}, \tau) \phi_S(-\vec{q},\tau)}_{\rm ret}
=\frac{1}{\beta g_S^2 N_S(0)}
\sum_{\omega_n}
\frac{1}{B \displaystyle{\frac{\omega_n^2}{k_F g_S |\bar \Phi|}}
+A \displaystyle{\frac{|\omega_n| \sqrt{\kappa_1 \kappa_2^{-1} q_x^4+q_y^2}}{k_F g_S|\bar \Phi|}}
+\kappa_1 q_x^4+ \kappa_2 q_y^2}
\label{eq:finite_T_smectic_phase}
\end{equation}
\end{widetext}
The most divergent term in $\chi^S_\perp(\vec q)$ has the asymptotic behavior $ T/( \kappa_1 q_x^4+\kappa_2 q_y^2)$, which implies that the Fourier transform of $\chi^S_\perp(\vec q)$, the transverse susceptibility of the smectic Goldstone mode,  is infrared divergent
at finite $T$. This divergence implies that the equal-time 
correlation functions of the smectic order parameter field $\langle \Phi(\vec x)^\dagger \Phi(\vec y) \rangle$ exactly vanishes, for all $\vec x \neq \vec y$. Thus, this system exhibits a form of  ``local quantum criticality'', {\it i.e.\/}
where the correlation length of the equal-time correlation function essentially
vanishes but the equal-position auto-correlation function scales as a function of time. In Fourier space, this means that the finite temperature dynamical susceptibility scales in frequency but not in momentum. This behavior was discussed recently in the context of the FL-nematic QCP\cite{lawler-2007} and in the quantum critical behavior of quantum dimer models\cite{Ghaemi05}. 

The 
vanishing of the equal-time correlation length however invalidates the assumption of
small smectic Goldstone fluctuations, where topological defects (dislocations) 
become important. With dislocations, the smectic order parameter vanishes (as translation invariance has been restored) and the equal-time correlation functions of the smectic order parameter become short ranged. This physics proceeds, as usual, by the non-perturbative Kosterlitz-Thouless mechanism of defect proliferation.
At low temperatures, where the density of the 
dislocations is low, the correlation length diverges, which recovers the 
proper behaviors of the zero-temperature quantum smectic phase.

For systems with a lattice background, the nature of the thermal nematic phase transition is determined by the point group symmetry of the lattice, {\it e.g.\/} for the square lattice it is an Ising transition. Also in the presence of a lattice there is a stable electronic smectic phase at finite temperature. In this case
the finite-temperature nematic-smectic transition is in the KT
universality class for an unpinned CDW. With a lattice background, the 2D unpinned smectic 
will be at its lower critical dimension. Therefore, the smectic phase has quasi long-range order
and a finite-temperature KT transition is expected. For smectics pinned by lattice,
the transition instead belongs to the universality class of the $q$-state Potts model, with $q$
equals to $n$ for a lattice with $n$-fold rotational symmetry \cite{Barath2007}. 
On the nematic side, Eq.
\eqref{eq:finite_T_smectic}
is still valid, but $\Delta \to 0$ as we approach the finite-temperature phase transition.
On the unpinned smectic side, the most singular term of the correlation function of its Goldstone mode will be $\sim T/q^2$,
which now will be unaffected by topological defects in the finite-$T$ smectic
phase since in the case of a lattice the dislocations are confined. This behavior of a phase smectic phase mode leads, as usual, to an 
equal-time correlation functions of the order parameter with power-law behavior as a function of distance. In the high-temperature phase, where the dislocations become deconfined, the order parameter has only short-ranged spatial correlations.

\section{Discussion and Conclusions}
\label{sec:conclusions}

In this paper we presented a phenomenological theory of quantum phase transitions in electronic liquid crystal phases. For simplicity we have only considered the charge channel and discussed the behavior of two charge-ordered phases: the electron nematic and the electron smectic. The latter phase has the same symmetries as a charge stripe phase and a unidirectional CDW. We have not discussed the behavior of spin excitations and the associated magnetic phases, {\it i.e.\/} a spin stripe.   The main results are summarized in Table \ref{table:summary}.

We discussed in detail how to describe a fluctuating charge stripe phase, a metallic electron nematic phase close to a quantum phase transition to an electronic smectic (or stripe) phase. We analyzed the nature of the fluctuations of the bosonic collective modes in the nematic as this quantum phase transition is approached: it is described by a low-energy Goldstone nematic mode (with dynamic critical exponent $z=3$) and a higher energy smectic collective mode (with dynamic critical exponent $z=2$), and discussed in detail the behavior of the nematic and smectic correlation functions in this regime. We also gave a detailed analysis of the electron smectic phase and of the behavior of the correlation functions in this phase. In particular we give explicit form of the dynamical susceptibilities at zero temperature, including the dynamics induced by the fermionic fluctuations. This analysis was also done in the electron smectic phase. The behavior of the nematic and smectic susceptibilities should be useful to interpret experiments that can probe this dynamics, particularly light-scattering experiments. With some minor changes the form of these correlation functions and susceptibilities also apply to the analysis of magnetic fluctuations, such as neutron scattering experiments in the  fluctuating stripe regime.

We also developed a description of the quantum critical behavior at this phase transition to an electronic smectic phase.
This effective critical theory is a quantum-mechanical generalization of the classical McMillan-deGennes theory for the nematic-smectic liquid crystals to quantum-mechanical metallic phases with the same pattern of symmetry breaking. It is a theory at zero temperature and it includes the effects of strong quantum fluctuations which turn out to have a very different character than their classical counterpart.  Due to the non-trivial effects of the fermionic fluctuations, this quantum phase transition is continuous whereas the classical finite-temperature transition is weakly first order. 

We presented an extensive discussion of the fate of the fermionic quasiparticles in each phase and at the quantum phase transition, and of the resulting non-FL behaviors. The resulting non-FL effects are quite rich. These results were obtained within a perturbative expansion in powers of the coupling between the fermions and the order-parameter fluctuations. As such, the obtained non-FL behaviors represent primarily a breakdown of perturbation theory rather than asymptotically exact results. The reliability of these low-order results will be checked in the near future using non-perturbative approaches such as higher dimensional bosonization. In any case our perturbative analysis of the quasiparticle self-energy shows once again that electronic liquid crystal phases are naturally compatible with non-FL behavior.

Although much of the theory that we presented is developed in the context of a continuum system, {\it i.e.\/} a system in which the effects of the coupling of the underlying lattice on the electronic order is ignored, we also gave a qualitative analysis of the symmetry-breaking effects resulting from the coupling to the lattice. This analysis will be generally correct even though the details of the microscopic band structures were ignored, as they will be reflected in the form of symmetry breaking fields and in their coupling constants. We expect that the discussion of the continuum (isotropic) theory will be applicable if the effects of the coupling to the lattice are comparatively weak (as expected far from van Hove singularities) and at temperatures high compared to the scale of these couplings. Quite surprisingly, although the pinning of the nematic Goldstone mode by the lattice was expected, we found that the smectic collective modes in general remain strongly fluctuating as the QCP is approached and into the smectic phase. This happens provided that the magnitude of the ordering wave vector obeys $Q_S<2k_F$.  Contrary to what happens in the case of the FL-nematic quantum phase transition, in which lattice effects often drive the transition first order, the nematic-smectic transition can naturally be continuous even in the presence of the coupling to the lattice.

We also considered the effects of low-temperature thermal fluctuations, and showed that the signatures of the electronic smectic phase can be detected through heat-capacity measurements as subleading corrections to FL. In the electronic nematic phase and at the nematic-smectic QCP, the heat capacity is 
dominated by nematic fluctuations, which goes as $T^{2/3}$. At the QCP,
the smectic mode yields an additional contribution $\sim T$ for $Q_S<2k_F$, a subleading term $\sim T^{3/2}$
for commensurate CDW with $Q_S=2k_F$ and $\sim T^{4/3}$ for the case of inflection points. In the electronic smectic phase, the smectic Goldstone mode gives a contribution to the heat capacity of $C\sim T^{3/2}$ in the absence of lattice symmetry breaking effects and $T^{2}$ with lattice symmetry breaking effects.

The approach that we followed is semi-phenomenological and it is based on the picture of a Fermi liquid that is coupled to an effective field theory describing the fluctuations of the nematic and smectic (stripe)  order parameters. This line of attack can be justified, at the level of mean-field theory, in weakly coupled systems based on the existence of a Fermi liquid for some range of parameters and its subsequent instabilities. Examples of this approach is the work on the FL/electron-nematic quantum phase transition of Oganesyan {\it et al} \cite{Oganesyan2001} in continuum models and of Metzner and coworkers\cite{Halboth-2000,Metzner-2003,yamase-2005,dellanna-2006} and H.-Y. Kee and coworkers\cite{kee-2003,Khavkine-2004} in lattice models. The extension of these works to the electron nematic/stripe phase transition that we discuss here is possible and we have obtained some unpublished results in this direction. However, this quantum phase transition requires that a coupling constant be larger than a critical value, which typically is not small, and hence the reliability of mean-field methods in this regime is at least problematic. This will be discussed in a separate publication. 

We have not discussed the realization of these electronic liquid crystal phases, and of the quantum phase transitions we discussed in microscopic models of strongly correlated systems. Nevertheless sufficiently close to a  continuous quantum phase transition this approach is likely to give the correct universal behavior. However, the use of theories based on the breakdown of the FL state is problematic in the strong correlation regime. In addition, so far there is no microscopic model in which both nematic and stripe phases are known to occur. The existence of an electron nematic phase in a microscopic model of a strongly correlated system has only been shown for the strong coupling regime of the Emery model of the cuprates in the asymptotically low doping regime\cite{kivelson04}. The existence of a metallic stripe phase in the same model for the doping range $x \sim 1/8$ has been suggested by a variational wave function (and hence mean-field in spirit) approach which projects out double occupancies of Cu sites \cite{lorenzana04,seibold-2004}. This suggests that the strongly coupled Emery model may more generally exhibit both nematic and stripe phases in its phase diagram. A number of publications have attempted to describe both stripe and nematic phases in Hubbard, $t-J$,  and Emery models using slave-particle methods \cite{yamase-2000,miyanga-yamase-2006,yamase-2006b}. However, in spite of their widespread use in the field, slave-particle mean-field theories are notoriously unreliable. 

It would be highly desirable to have high quality numerical simulations to address this problem in models of strongly correlated systems. Density matrix renormalization group (DMRG) calculations have provided strong evidence for stripe correlations in Hubbard-type models on narrow strips (with up to 5 legs) \cite{white-2000,carlson04}. However, the geometry used in DMRG, which breaks the rotational invariance under $90^\circ$ rotations of the square lattice explicitly makes it difficult to distinguish a stripe from a nematic phase. The same problem arises in finite-size diagonalizations of small systems.  Quantum Monte Carlo simulations are less affected by such geometric limitations but suffer from the notorious fermion sign problem at low temperatures. QMC simulations have indeed shown an increase in nematic fluctuations at low temperatures in Hubbard type models (see a discussion in Ref. \cite{carlson04}) but, as far as we know, not yet in the Emery model.

\begin{acknowledgments}
We are grateful to Steve Kivelson for many useful comments and criticisms, and to Andrei Chubukov for discussions.
This work was supported in part by the
National Science Foundation
grants DMR 0442537 and DMR 0758462 at the University of Illinois (EF), and  by the Department of Energy, 
 Division of Basic Energy Sciences
under Award DE-FG02-07ER46453 through the Frederick Seitz Materials
Research Laboratory at UIUC (KS, BMF, EF),  the Stanford Institute of Theoretical Physics (EF), and by 
NSERC, CIFAR and CRC (MJL).
\end{acknowledgments}

\appendix
\section {Tensor Form of the Order Parameters}
\label{app:sec:tensor}

In this appendix we rewrite the order parameters in a tensorial form which makes their correct symmetry transformation properties apparent. Much of what is done here follows closely the analysis of the classical case\cite{Chaikin98}.

In 2D, because the rotation group $SO(2)$ group is Abelian, and has only two 
$1D$ representations with l=2 ($l_z=\pm2$), it follows that the nematic order parameter {\bf N}, a $2\times 2$ symmetric traceless tensor, has just two 
independent components, $n_{11}$ and $n_{12}$, as shown in Eq. 
\eqref{eq:2D_nematic}.
The Abelian nature of the $SO(2)$ group enables us to use complex 
numbers, instead of tensors to represent the action of the 
group, as shown in Eq. \eqref{eq:2D_nematic_complex}.

All the formulas we have presented in the main text using the complex order 
parameter can be translated into the tensor language, which can more naturally be generalized
to higher dimensions. Thus, Eq. \eqref{eq:frank_term_complex} becomes
\begin{align}
-2 \kappa \; \tr\left\{{\bf N}(\vec{r})\left[{\bf D}{\bf N}(\vec{r})
{\bf N}(\vec{r})\right]\right\}.
\label{eq:frank_term_tensor}
\end{align}
${\bf D}$ is the rank $2$ tensor formed by the outer product of the two-component real
vector $(\partial_x,\partial_y)$:
\begin{align}
{\bf D}=
\left(\begin{array}{c}
\partial_x
\\
\partial_y
\end{array}
\right) \otimes 
\left(\partial_x,\partial_y\right),
\end{align}
In Eq \eqref{eq:frank_term_tensor} the first derivative operator acts on the second ${\bf N}$ factor, while the second 
derivative operator acts on the last ${\bf N}$ factor. Here only the traceless part of 
${\bf D}$, ${\bf D}-\tr[{\bf D}] {\bf I}/2$, gives non-zero contribution to Eq. \eqref{eq:frank_term_tensor}. The assignment of 
derivatives is not unique. But only this term gives a linearly independent 
contribution to the nematic Goldstone mode. Other assignments either have no 
contribution, or just give a contribution proportional to this one. Therefore, 
there is no needs to study other terms. 

Similarly, Eq. \eqref{eq:s_n_coupling_complex} has the tensor 
form:
\begin{widetext}
\begin{equation}
S_{\textrm{int}}= -2 g \int \frac{\mathrm{d}\vec{k}\mathrm{d}\Omega}{(2\pi)^3}
\int \frac{\mathrm{d}\vec{q}\mathrm{d}\omega}{(2\pi)^3}
\frac{1}{k^2}
\tr\left\{ {\bf N(\vec{q},\omega)}
\left(\begin{array}{cc}
k_{x}^2 & k_{x} k_{y}
\\
k_{x} k_{y} & k_{y}^2
\end{array}
\right)\right\}
\rho(\vec{k}-\vec{q}/2,\Omega-\omega/2) 
\rho (-\vec{k}-\vec{q}/2,-\Omega-\omega/2).
\label{eq:s_n_coupling_tensor}
\end{equation}
\end{widetext}
Again, it is easy to check that only the traceless part of the tensor composed by the momenta
$\vec{k}$ is needed here, since the trace of that tensor has no contribution. Notice that although there is
a factor of $1/k^2$ in our formula, it will not cause divergence, 
since what we are interested in is  a CDW, a state that orders at a finite wave vector $k\sim Q_S$.

Finally, using the fermionic density quadrupole tensor defined in Eq. 
\eqref{eq:density_quadrupole}, Eq. \eqref{eq:g_N} can be written in tensorial form as
\begin{align}
-g_N \int\mathrm{d}\vec{r}\mathrm{d}t \; \tr[\mathbf{Q}\mathbf{N}]
\label{eq:g_N_complex}
\end{align}

\section{Quantum Critical Point for $Q_S<2k_F$}
\label{app:sec:critical}

As discussed in the main text, at the nematic-smectic critical point 
$Q_S<2 k_F$, the nematic Goldstone mode is the low-energy mode. To obtain its effective low-energy effective action we integrate the high energy mode of the smectic field, $\Phi$. 
Terms like
\begin{align}
\langle  (\partial_x \Phi^\dagger \Phi) (\Phi^\dagger \partial_x \Phi) \rangle \phi_N \phi_N,
\end{align}
will generate the leading corrections to the $\phi_N$ propagator. 
We evaluate the integral numerically to deduce the kernel for the quadratic fluctuations.
To the one-loop level, it is
\begin{align}
C_1 q^2+i C_2 \omega^3 |\log(\omega/q^2)|/q^4,
\end{align}
for $0<\omega\ll q^2$.
The first term $C_1 q^2$ renormalizes the two Frank 
constant $K_1$ and $K_2$, while the second one is subleading compared to the dynamical term of the nematic Goldstone mode, which is proportional to $i|\omega|/q$. This result suggests that the smectic field $\Phi$
is an irrelevant perturbation to the nematic Goldstone mode at the nematic-smectic critical point.

Terms of the form
\begin{align}
\langle \partial_x \Phi^\dagger \phi_N \partial_x \Phi \phi_N \rangle \Phi^\dagger \Phi
\end{align}
provide leading order corrections to the propagator of the smectic field $\Phi$,
and yield a self-energy correction to the field $\Phi$. The loop integral is computed
numerically and is well fit by the form
\begin{align}
\sim C_3 q^2+i C_4 \omega^{5/4},
\end{align}
when $\omega \gg q^3$.
The first term renormalizes the constants $C_x$ and $C_y$ defined in
Eq. \eqref{eq:ns_action} and the second one is subleading compared to the dynamical term of the smectic field which is linear in $|\omega|$.

Therefore, these arguments provide strong evidence that the coupling between the nematic Goldstone mode $\phi_N$ and the smectic order parameter $\Phi$ is
irrelevant. In the low-energy theory, we can treat in practice $\phi_N$ and $\Phi$ as two separate
modes.

\section{Non-analytic Terms of the Effective Field Theory of the Electron Smectic with $Q_S=2k_F$}
\label{app:sec:phi_nonanalytic}

\begin{widetext}
The mean-field Hamiltonian of the smectic phase is
\begin{align}
H_{MF}=\int\!\frac{d^2 k}{(2\pi)^2}\!
	\left(\epsilon(\vec{k}) \psi^\dagger(\vec{k}) \psi(\vec{k})
	      +g_S \; \bar \Phi \; \psi^\dagger(\vec{k}\!+\!\vec{Q}_S)\psi(\vec{k})
	+g_S \; \bar \Phi^* \; \psi^\dagger(\vec{k}\!-\!\vec{Q}_S)\psi(\vec{k})\right)
	+\Delta_S |\bar \Phi|^2.
\end{align}
The smectic order parameter $\bar \Phi$ will act as a periodic background potential 
in the direction of $\vec{Q}_S$ which will reconstruct the band structure and the FS. For $Q_S=2k_F$, we will only consider
the lowest two bands. By ignoring higher bands, we can diagonalize this mean-field 
Hamiltonian to get the single particle dispersion relation
\begin{align}
E_{\pm}(\vec{k}) =\frac{1}{2}\left(\epsilon(\vec{k}) + \epsilon(\vec{k}+\vec{Q}_S)\right)
\pm 
\sqrt{\frac{1}{4}\left(\epsilon(\vec{k}) - \epsilon(\vec{k}+\vec{Q}_S)\right)^2 + g_S^2|\bar \Phi|^2 }.
\end{align}

\end{widetext}
Here the $+$ sign is for the upper band and the $-$ sign is for the lower
band. The lower band is partially filled by the fermions, but the
upper band is empty. So the Landau free energy will be
\begin{align}
	F(|\bar \Phi|)= \int_{E_{-}(\vec{k})<\mu} \frac{d^2 k}{(2\pi)^2} E_{-}(\vec{k})
	+\Delta_S |\bar \Phi|^2,
\end{align}
where $\mu$ is the chemical potential.
By expanding the dispersion relation around the two points marked on Fig. 
\ref{fig:FSCDW}(b)
\begin{align}
	\epsilon_{1,2}(\vec{q})= \pm q_y+ \frac{\kappa}{2} q_x^2,
\label{eq:dispersion_q_2kf}
\end{align}
where $\vec{q}$ is measured from the special points marked in 
Fig. \ref{fig:FSCDW}(b). The Fermi velocities at
these two points are just opposite to each other, $v_F=\pm 1$, and the curvatures 
of the FS have the same value $\kappa$ at these two points.

By using this approximate dispersion relation, the Landau free energy
can be determined as
\begin{align}
	F(|\bar \Phi|)=F(0)+r |\bar \Phi|^2+u' |\bar \Phi|^{5/2}.
\end{align}
Here the coefficient of the quadratic term $r$ depends on the high-energy 
cutoff, which reflects that the Landau free energy depends on the band structure 
all the way down to the bottom of the band. 

On the other hand, $u'$ is universal, as it depends only on the dispersion 
relation around the two special points,
\begin{align}
	u'=-\frac{\Gamma(-1/4)}{5\pi^{3/2}\Gamma(1/4)}
	\frac{g_s^{5/2}}{\sqrt{\kappa/2}}
	\approx 0.049\frac{g_S^{5/2}}{\sqrt{\kappa/2}}.
\end{align}
$u'$ diverges in the limit of $\kappa\rightarrow 0$, which means
terms in lower order, such as $|\bar \Phi|^9/4$, will be generated (see below).

Around an inflection point, the dispersion relation is shown in Eq. 
\eqref{eq:inf_disp}. Similar to the calculation above, a term proportional to
$|\bar\Phi|^{9/4}$ will be found. The coefficient of these term $u''$
will be
\begin{align}
	u''=-\frac{2\Gamma(-1/8)}{9\pi^{3/2}\Gamma(3/8)}
	\frac{g_S^{9/4}}{b^{1/4}}
	\approx 0.15\frac{g_S^{9/4}}{b^{1/4}}
\end{align}

\section{Goldstone Mode in the Smectic Phase}
\label{app:sec:gs_mode_smectic}

In the smectic phase, the smectic order parameter obtains an expectation value. 
We write $\bar \Phi$ as
\begin{align}
\Phi=\bar \Phi+\delta \Phi.
\end{align}
$\bar \Phi$ can be considered as a periodic potential background for the 
fermions. As a result, the fermions will form band structure in the $y$ direction. 
Define the Bloch states as
\begin{align}
\psi_n (\vec{k})=\sum_m T_{n,m}(\vec{k}) e^{i m\phi} \psi(\vec{k}+m \vec{Q}_S),
\end{align}
where $T_{n,m}$ is an orthogonal transfer matrix which depends on the amplitude of
the order parameter, and $\phi$ is the phase of $\bar \Phi$. In general, the Bloch 
wave is related to the plane wave by a unitary transformation. Here,
due to the fact that only one harmonic of the CDW with wave vector $\vec{Q}_S$ is considered, after a proper spatial 
translation, which is a shift of $\phi$, it can be simplified to an 
orthogonal transformation, which enables us to define the orthogonal transfer 
matrix $T_{n,m}$. The inverse formula can also be written down, since the 
$T_{n,m}e^{i m\phi}$ must be unitary: 
\begin{align}
\psi (\vec{k}+m \vec{Q}_S)=\sum_n T_{n,m}(\vec{k}) e^{-i m\phi} \psi_n (\vec{k}).
\label{eq:transformation}
\end{align}

Integrating out the fermionic degrees of freedom, the leading term in the power 
series of the smectic fluctuations $\delta \Phi$ starts from the quadratic order.
\begin{widetext}
\begin{align}
\mathcal{L}=-\left(\begin{array}{cc}
\delta \Phi^{\dagger}(\vec{q},\omega) & \delta \Phi(-\vec{q},-\omega)	
\end{array}\right)
\left(\begin{array}{cc}
\Delta_{S}-\Pi(\vec{Q}_S+\vec{q},\omega) & -\Pi'(\vec{Q}_S+\vec{q},\omega)
\\
-\Pi'(-\vec{Q}_S-\vec{q},-\omega) & \Delta_{S}-\Pi(\vec{Q}_S-\vec{q},-\omega)
\end{array}\right)
\left(\begin{array}{c}
\delta \Phi(\vec{q},\omega)
\\
\delta \Phi^{\dagger} (-\vec{q},-\omega)	
\end{array}\right).
\label{eq:stripefluctuations}
\end{align}
Notice that $\delta \Phi^{\dagger}(\vec{q},\omega)\neq \delta \Phi(-\vec{q},-\omega)$, since the field $\Phi$ is complex. The continuous translational symmetries have been broken into discrete ones. 
Therefore, a process which changes the momentum by $n \vec{Q}_S$, 
where $n$ is an integer, is allowed. This is the reason why we have 
the terms such as 
$\delta \Phi^{\dagger} (\vec{q},\omega)\delta \Phi^{\dagger} (-\vec{q},-\omega)$.

$\Pi(\vec{Q}_S+\vec{q},\omega)$ in the diagonal terms is the standard fermion 
bubble integral in the smectic phase but with momentum close to $\vec{Q}_S$,
\begin{align}
&\Pi(\vec{Q}_S+\vec{q},\omega)=
\int \frac{\mathrm{d} \vec{k}_1\mathrm{d} \Omega_1}{(2\pi)^3}
\int \frac{\mathrm{d} \vec{k}_2\mathrm{d} \Omega_2}{(2\pi)^3}
\nn\\
&\left\langle
\psi^{\dagger}(\vec{k}_1+\vec{Q}_S+\vec{q}/2,\Omega_1+\omega/2)
\psi(\vec{k}_1-\vec{q}/2,\Omega_1-\omega/2) 
\psi^{\dagger}(\vec{k}_2-\vec{q}/2,\Omega_2-\omega/2)
\psi(\vec{k}_2+\vec{Q}_S+\vec{q}/2,\Omega_2+\omega/2)
\right\rangle.
\nn\\
\end{align}
Since the eigenstates are Bloch waves, we need to transfer $\psi$ into $\psi_n$. 
\begin{align}
&\Pi(\vec{Q}_S+\vec{q},\omega)
\nn\\
&=\sum_{m_1,m_2,n_1,n_2}
\int \frac{\mathrm{d} \vec{k}}{(2\pi)^2}
T_{n_1,m_1+1}(\vec{k}+\vec{q}/2) T_{n_2,m_1}(\vec{k}-\vec{q}/2)
T_{n_2,m_2}(\vec{k}-\vec{q}/2) T_{n_1,m_2+1}(\vec{k}+\vec{q}/2) 
F_{n_1,n_2}(\vec{k},\vec{q},\omega)
\end{align}
where $F_{n_1,n_2}(\vec{k},\vec{q},\omega)$ describe the scattering between 
fermions in band $n_1$ and $n_2$, which is defined as
\begin{align}
-\int \frac{\mathrm{d} \Omega}{2\pi}
\left \langle
\psi_{n_1}^{\dagger}\left(\vec{k}+\vec{q}/2,\Omega+\omega/2\right)
\psi_{n_1}\left(\vec{k}+\vec{q}/2,\Omega+\omega/2\right)
\right\rangle
\left\langle
\psi_{n_2}^{\dagger}\left(\vec{k}-\vec{q}/2,\Omega-\omega/2\right)
\psi_{n_2}\left(\vec{k}-\vec{q}/2,\Omega-\omega/2\right) 
\right\rangle.
\end{align}

The off-diagonal $\Pi'(\vec{Q}_S+\vec{q},\omega)$ is very similar to 
$\Pi(\vec{Q}_S+\vec{q},\omega)$.
\begin{align}
&\Pi'(\vec{Q}_S+\vec{q},\omega)=
\int \frac{\mathrm{d} \vec{k}_1\mathrm{d} \Omega_1}{(2\pi)^3}
\int \frac{\mathrm{d} \vec{k}_2\mathrm{d} \Omega_2}{(2\pi)^3}
\nn\\
&\left\langle
\psi^{\dagger}(\vec{k}_1+\vec{Q}_S+\vec{q}/2,\Omega_1+\omega/2)\psi(\vec{k}_1-\vec{q}/2,\Omega_1-\omega/2) 
\psi^{\dagger}(\vec{k}_2+\vec{Q}_S-\vec{q}/2,\Omega_2-\omega/2)\psi(\vec{k}_2+\vec{q}/2,\Omega_2+\omega/2)
\right\rangle.
\nn\\
\end{align}
Again, we need to transfer $\psi$ into $\psi_n$.
\begin{align}
&\Pi'(\vec{Q}_S+\vec{q},\omega)
\nn\\
&=e^{-2 i \phi}\sum_{m_1,m_2,n_1,n_2}
\int \frac{\mathrm{d} \vec{k}}{(2\pi)^2}
T_{n_1,m_1+1}(\vec{k}+\vec{q}/2) T_{n_2,m_1}(\vec{k}-\vec{q}/2)
T_{n_2,m_2+1}(\vec{k}-\vec{q}/2) T_{n_1,m_2}(\vec{k}+\vec{q}/2) 
F_{n_1,n_2}(\vec{k},\vec{q},\omega)
\nn\\
\end{align}
\end{widetext}

To study the low-energy excitations, we need to diagonalize the matrix of 
Eq. \eqref{eq:stripefluctuations}. Then two eigenmodes appear. In 
the limit where we take $\omega \rightarrow 0$ first, then $q \rightarrow 0$, 
it is easy to see that eigenvalues of these two eigenmodes are 
$\Pi(\vec{Q}_S,0)+\Delta_S \pm |\Pi'(\vec{Q}_S,0)|$.
The equation
\begin{align}
\Pi(\vec{Q}_S,0)+\Delta_S-\Pi'(\vec{Q}_S,0)=0
\end{align}
reproduces the mean-field self-consistent equation which determines $\bar \Phi$. 
Therefore, the mode which takes the minus sign will be the Goldstone mode
and the other is the amplitude mode. 

As for the Goldstone mode, the intra-band scattering $F_{n,n}$ and the 
inter-band scattering $F_{n_1,n_2}$ where $n_1\ne n_2$, have different 
contributions. The intra-band scattering have no contribution in the 
$q\rightarrow 0$ limit, because its contribution to $\Pi(\vec{Q}_S,0)$ 
and $|\Pi'(\vec{Q}_S,0)|$ cancels.

When $n_1=n_2$, term in $\Pi(\vec{Q}_S,0)$ is
\begin{widetext}
\begin{align}
\int \sum_{m_1,m_2,n}\frac{\mathrm{d} \vec{k}}{(2\pi)^2}
T_{n,m_1+1}(\vec{k})T_{n,m_1}(\vec{k}) T_{n,m_2}(\vec{k}) T_{n,m_2+1}(\vec{k})
F_n(\vec{k},0,0),
\end{align}
and term in $\Pi'(\vec{Q}_S,0)$ is
\begin{align}
e^{-2 i \phi}\int \sum_{m_1,m_2,n}\frac{\mathrm{d} \vec{k}}{(2\pi)^2}
T_{n,m_1+1}(\vec{k}) T_{n,m_1}(\vec{k})
T_{n,m_2+1}(\vec{k}) T_{n,m_2}(\vec{k}) 
F_n(\vec{k},0,0).
\end{align}
\end{widetext}
From these two equations, the contributions are canceled for the Goldstone mode.

The inter-band scatterings are gapped, so that we have
\begin{align}
F_{n_1,n_2}(\vec{k},0,0)=
-\frac{n_f\left(E_{n_1}(\vec{k})\right)
-n_f\left(E_{n_2}(\vec{k})\right)}
{E_{n_1}(\vec{k})-E_{n_2}(\vec{k})},
\end{align}
where $E_n(\vec{k})$ is the eigen-energy of fermions in band $n$ 
with momentum $\vec{k}$ and $n_f$ is the Fermi distribution function. 
Summing all the inter-band contributions, the equation $\Pi(\vec{Q}_S,0)+\Delta_S-\Pi'(\vec{Q}_S,0)=0$ 
reproduces the mean-field self-consistent equation which determines 
$\bar \Phi$ as required by the Ward identity

Now let us study the frequency dependence of the smectic Goldstone mode. 
It is straight forward to show that the frequency dependence from 
the inter-band scatterings will start from the order $\omega^2$ 
because the inter-band scattering has an energy gap.

We expect the intra-band scattering gives a dynamical term $\sim i |\omega|/q$
to $\Pi$ and $\Pi'$ since there is no energy gap and the $T$ matrix has no 
singular points. However, as we just showed, the intra-band scattering
contributions to $\Pi$ and $\Pi'$ cancels at $q=0$. Hence,
the leading dynamical term is $\sim i |\omega| q$. The Lagrangian density reads
\begin{widetext}

\begin{equation}
\mathcal{L}=g_S^2|\bar{\Phi}|^2 N_S(0) \left(B(\varphi_q)\frac{\omega^2}{k_F g_S |\bar \Phi|}+i A(\varphi_q) 
\frac{|\omega| q}{k_F g_S |\bar \Phi|}-\kappa_S(\varphi_q) q^2 \right) |\phi_\Phi(\vec q,\omega)|^2,
\end{equation}

\end{widetext}

Here $\bar \Phi$ is the expectation value of the smectic order parameter
and the smectic Goldstone mode $\phi_\Phi$ is defined in Eq. \eqref{eq:phi_fluctuation}.
$\varphi_q$ is the angle between $\vec{q}$ and the stripe direction and $N_S(0)$
is the density of states in the smectic phase. $A(\varphi_q)$, $B(\varphi_q)$ and 
$\kappa_S(\varphi_q)$ are coupling constants that depend on microscopic details and 
the direction of $\vec{q}$ which reflects the anisotropic nature of the smectic 
phase. 

This low-energy theory has $z=1$ which is different from $z=2$ at the QCP. Notice 
that the denominator of the first two terms in the above formula contains
$|\bar \Phi|$, which just means that the $z=1$ breaks down when on approach the QCP
from the smectic side. On the other side, if we notice that there is a 
coefficient $|\bar{\Phi}|^2$ in the front of the Lagrangian density, the $|\bar \Phi|$
in the denominator cause no divergence at small $|\bar \Phi|$. In fact, as we
can see that the intra-band scattering vanishes as $|\bar{\Phi}|$ goes to $0$, which
is what one should expected.

It is easy to check that the coefficients of $A(\varphi_q)$ and $B(\varphi_q)$ are dimensionless, 
while $B(\varphi_q)$ and $\kappa_S(\varphi_q)$ are real. $A(\varphi_q)$ is a real function for most
values of $\varphi_q$. However, from some special cases, for example Fig. \ref{fig:FSCDW}(d),
there will be no FS whose Fermi velocity is close to the $y$ direction. As a result, the
particle-hole excitations with momentum close to the $x$ direction will not be damped.
Therefore, $A(\varphi_q)$ will be imaginary for $\varphi_q\sim 0$ or $\pi$.

\vskip 1cm

\section{RPA Calculation of the Fermion Self-energy}
\label{app:sec:RPA}

The imaginary part of the fermion self-energy corrections at the FS, $\Sigma^{\prime \prime}(\vec{k}_F,\omega)$, scales with frequency (as $\omega \to 0$) as $\sim |\omega|^\mu$. When the scaling index $\mu$ is larger than $1$,
the low-energy theory of the fermions can be described by the Landau FL 
theory. But if $\mu$ is less than $1$, the self-energy correction will dominate 
and the Landau FL theory 
will break down at low energies. As a result, the system will  exhibit non-FL behavior.

\subsection{The nematic-smectic QCP}
\label{app:sub:sec:n_s_critical_point}

To the one-loop level, for $\omega>0$,the imaginary part of the fermion self-energy correction 
from $\Phi$ is
\begin{widetext}

\begin{equation}
\Sigma^{\prime \prime}_S(\vec{k}_F,\omega)=\displaystyle{\frac{g_S^2}{2}}
\int_{0<\epsilon(\vec{k}_F+\vec{Q}_S-\vec{q})<\omega}
\displaystyle{\frac{\mathrm{d}\vec{q}}{(2\pi)^2} }\;
B_{S}\left(\vec{q},\omega-\epsilon\left(\vec{k}_F+\vec{Q}_S-\vec{q}\right)\right),
\label{eq:damping_s}
\end{equation}
\end{widetext}
where $B_{S}(\vec{q},\omega)$ is the spectral density of the smectic field. 
At the nematic-smectic QCP ($Q_S<2 k_F$) Eq. 
\eqref{eq:smectic-susceptibility} gives
\begin{align}
B_{S}\left(\vec{q},\omega \right)=\displaystyle{\frac{2 C_0 \omega}{C_0^2\omega^2+(C_x q_x^2+C_y q_y^2)^2}}.
\end{align}

By substituting this to Eq. \eqref{eq:damping_s}, we find that for the four points 
marked on Fig. \ref{fig:FSCDW}(a), $\Sigma^{\prime \prime}_S \sim |\omega|^{1/2}$, while for all other 
points on the FS $\Sigma^{\prime \prime}_S \sim \omega^2$. For the special case of $C_x=C_y=C$, an analytical 
formula can be achieved. For the four special points, 
$\Sigma^{\prime \prime}_S=\sqrt{2|\omega|}/(4\pi\sqrt{C C_0})$ 
valid for $|\omega| C/C_0 \ll 1$. For other points on the FS, 
$\Sigma^{\prime \prime}_S(\vec{k}_F)=\omega^2 v^2 C_0 /(16\pi C^2 \Delta^3)$, where the constants $\Delta$
and $v$ comes from the expansion of the dispersion relation near $\vec{k}_F+\vec{Q}_S$
as $\epsilon(\vec{q}+\vec{k}_F+\vec{Q}_S)=\mu+\Delta+\vec{v}\cdot\vec{q}+\ldots$.

The nematic Goldstone mode has an $\omega^{2/3}$ contribution 
to $\Sigma^{\prime \prime}$ for
most points on the FS except the four points in the main-axis directions where $\Sigma^{\prime \prime}\sim|\omega|^{3/2}$ \cite{Oganesyan2001}. The 
$|\omega|^{2/3}$ behavior  is dominant over a $\Sigma^{\prime \prime}_S \sim \omega^2$ scaling, but not over
a $\Sigma^{\prime \prime}_S \sim |\omega|^{1/2}$ contribution. As a result, the four main-axis directions still have
FL behavior. The special points shown in Fig. \ref{fig:FSCDW}(a) will have 
$\Sigma^{\prime \prime} \sim |\omega|^{1/2}$ non-FL behaviors. For all other part of the Fermi
surface, $\Sigma^{\prime \prime} \sim |\omega|^{2/3}$ due to the nematic Goldstone mode. If the nematic
Goldstone mode is gapped due to a lattice background or an external field, the 
$\Sigma^{\prime \prime}\sim|\omega|^{2/3}$ non-FL behavior will disappear. But the 
$\Sigma^{\prime \prime} \sim |\omega|^{1/2}$ non-FL behavior at the four special points will persist.

The incommensurate CDW of $Q_S=2k_F$ case has a relevant perturbation $\Phi^{5/2}$, so 
the critical theory will be controlled by microscopic details. RPA calculation will not 
be reliable in such a situation. 

However we know that the commensurate CDW with $Q_S=2k_F$ is 
a second order transition described by the Gaussian theory, so RPA is applicable. 
Here we follow the same approach  mentioned above.
For most part of the FS, this bosonic mode will provide a self-energy correction 
$\Sigma^{\prime \prime}_S\sim \omega^{2}$ for small $\omega$. However for the special point 
$\vec{k}=\pm k_F \vec{e}_y$, $\Sigma^{\prime \prime}_S$ is linear in 
$\omega$ when $\omega \kappa \ll 1$. 
$|\omega|^{2/3}$ term from the $\phi_N$ mode will not be present for the commensurate CDW due to the lattice background. Hence, the fermions at this critical point are described by a FL theory, except for fermions 
with $\vec{v}_F\parallel\vec{Q}_S$.

For the special case of the inflection points, by following the same procedures, 
we get $\Sigma^{\prime \prime}\sim |\omega|^{13/12}$ at inflection points, where the 
FL picture is valid for the whole FS.

\subsection{The electron smectic phase}
\label{app:sub:sec:smectic_phase}

In the smectic phase, for $\omega>0$, the fluctuations of the smectic Goldstone boson $\phi_\Phi$ will contribute to the fermion self-energy:
\begin{widetext}
\begin{equation}
\Sigma^{\prime \prime}_{n,\phi_\Phi}(\vec{k},\omega)=\sum_{n'}\frac{|\bar\Phi|^2}{2}
\int_{0<\epsilon_{n'}(\vec{k}-\vec{q})<\omega}\frac{\mathrm{d}\vec{q}}{(2\pi)^2} 
\left|\tilde{g}^{n,n'}_S(\vec{k},\vec{q})\right|^2
B_{\phi_\Phi}\left[\vec{q},|\omega|-\epsilon_{n'}(\vec{k}-\vec{q})\right].
\label{eq:damping}
\end{equation}
Here $B_{\phi_\Phi}(\vec{q},\omega)$ is the spectral density function of the 
$\phi_\Phi$ mode presented below, $\epsilon_{n}(\vec{k})$ is the dispersion relation for fermions in 
band $n$ and the vertex is 
\begin{align}
\tilde{g}^{n,n'}_S(\vec{k},\vec{q})=&i g_S \sum_m \left[T_{n,m+1}(\vec{k}-\vec{q})T_{n',m}(\vec{k})-
T_{n,m}(\vec{k}-\vec{q})T_{n',m+1}(\vec{k})\right].
\nonumber
\end{align}
with $T_{n,m}$ being the matrix element of the orthogonal transformation defined in Eq.~\eqref{eq:transformation}

The energy gap between two different energy bands dictated that the contributions from the interband scatterings to $\Sigma^{\prime \prime}$ ($n\ne n'$) scale as $\omega^2$ at small $\omega$, which is subleading to the FL behavior. As for the intraband scatterings, the vertex vanishes linearly in the long wavelength limit ($q\sim 0$) as
\begin{align}
\tilde{g}^{n,n}_S(\vec{k},\vec{q})
=&i \sum_m \left[\nabla T_{n,m}(\vec{k})T_{n,m+1}(\vec{k})
-\nabla T_{n,m+1}(\vec{k})T_{n,m}(\vec{k})\right]\cdot \vec{q}
+O(q^2).
\end{align}
This is because the Goldstone mode cannot couple directly to the fermion density but to its fluctuations, as required by the translational symmetry. As a consequence of this structure of the vertex, we found that the quasiparticle scattering rate from intraband scatterings are also subleading corrections to the FL behavior by numerically evaluating the integral in Eq. \eqref{eq:damping}.

For an electron  smectic without a lattice background, Eq. \eqref{eq:smectic_CDW_no_lattice} gives
\begin{equation}
B_{\phi_\Phi}(\vec{q},\omega)=\displaystyle{\frac{k_F}{g_S|\bar{\Phi}| N_S(0)}}\;
\displaystyle{\frac{2 A \omega \sqrt{\kappa_1\kappa_2^{-1} q_x^4+q_y^2}}
{\left(A \omega \sqrt{\kappa_1\kappa_2^{-1} q_x^4+q_y^2}\right)^2
+\left(B\omega^2- k_F g_S|\bar \Phi| (\kappa_1 q_x^4+ \kappa_2q_y^2)\right)^2}}.
\end{equation}
For most part of the FS, the intraband scatterings lead to a fermion scattering rate $\Sigma^{\prime \prime}_{\phi_\Phi} \sim \omega^2 \log |\omega|$ in the small frequency limit. However for some special points, where the Fermi velocity is perpendicular to the stripe direction,  $\Sigma^{\prime \prime}_{\phi_\Phi} \sim|\omega|^{3/2}$. This result suggests that the fermions will be described by a FL.
For unpinned smectics with a lattice background, we have 
\begin{align}
B_{\phi_\Phi}(\vec{q},\omega)=\displaystyle{\frac{k_F}{g_S|\bar{\Phi}| N_S(0)}}\;
\displaystyle{\frac{2 A(\phi_q) \omega q}
{\left(A(\phi_q) \omega q\right)^2
+\left(B(\phi_q) \omega^2- k_F g_S|\bar \Phi| \kappa_S(\phi_q)q^2\right)^2}},
\end{align}
\end{widetext}
and  that $\lim_{\omega \to 0} \Sigma^{\prime \prime} (\omega)\sim\omega^{2} \log |\omega|$ at most part of the Fermi surface, while  $\Sigma^{\prime \prime}(k_F,\omega)\sim\omega^2$ at the special points where $\tilde{g}^{n,n}_S(\vec{k},\vec{q})$ is independent of $q_t$ to the linear order of $\vec{q}$, with $q_t$ being the component of $\vec{q}$ parallel to the reconstructed Fermi surface.


\begin{thebibliography}{158}
\bibitem[{chi()}]{chiral}
\bibinfo{note}{{The term $i N^\dagger (\partial_x+i \partial_y) N^\dagger
  (\partial_x+i \partial_y) N+h.c.$ is not allowed due to chiral symmetry.}}
\end{thebibliography}
\end{document}